\documentclass[fleqn,oldversion]{aa}
\usepackage{natbib}
\bibpunct{(}{)}{;}{a}{}{,}
\usepackage{graphicx}
\usepackage{natbib}
\usepackage{bm}
\usepackage{amsmath}
\usepackage{amsfonts}
\usepackage{amssymb}
\usepackage{txfonts}

\newcommand{\z}{\phantom{0}}
\newcommand{\zz}{\phantom{00}}
\renewcommand{\a}{\phantom{^*}}
\newcommand{\pd}{\partial}

\newcommand{\phicapbar}{\,\overline{\!\Phi\!}\,}
\newcommand{\rhobar}{\,\overline{\!\rho}}

\begin{document}

\title{Exploring the relativistic regime with Newtonian hydrodynamics: \\
  II. An effective gravitational potential for rapid rotation}

\author{B.~M{\"u}ller \inst{1} \and
        H.~Dimmelmeier \inst{1,2} \and 
        E.~M{\"u}ller \inst{1}}

\offprints{Ewald M{\"u}ller, \\ \email{emueller@mpa-garching.mpg.de}}

\institute{Max-Planck-Institut f\"ur Astrophysik,
  Karl-Schwarzschild-Str.~1, 85741 Garching, Germany
  \and
  Department of Physics, Aristotle University of Thessaloniki,
  54124 Thessaloniki, Greece}

\date{Received date / Accepted date}


\abstract{We present the generalization of a recently introduced
  modified gravitational potential for self-gravitating fluids. The
  use of this potential allows for an accurate approximation of
  general relativistic effects in an otherwise Newtonian hydrodynamics
  code also in cases of rapid rotation. We test this approach in
  numerical simulations of astrophysical scenarios related to compact
  stars, like supernova core collapse with both a simplified and
  detailed microphysical description of matter, and rotating neutron
  stars in equilibrium. We assess the quality of the new potential,
  and demonstrate that it provides a significant improvement compared
  to previous formulations for such potentials. Newtonian simulations
  of compact objects employing such an effective relativistic
  potential predict inaccurate pulsation frequencies despite the
  excellent agreement of the collapse dynamics and structure of the
  compact objects with general relativistic results. We analyze and
  discuss the reason for this behavior.}

  \keywords{gravitation -- hydrodynamics -- methods: numerical --
    relativity -- stars:supernovae:general}

\authorrunning{B.~M{\"u}ller et al.}
\titlerunning{An effective gravitational potential for rapid rotation}

\maketitle


\section{Introduction}
\label{sec:introduction}

Rapid rotation plays an important role in numerical simulations of a
number of astrophysical events connected to stellar core collapse and
neutron star evolution, such as accretion-induced collapse
\citep{dessart_06_a, dessart_07_a}, the collapsar scenario for long
gamma-ray bursts \citep{macfadyen_99_a, dessart_07_b}, the
phase-transition-induced collapse of a rotating neutron star to a
quark star \citep{lin_06_a}, or rotational instabilities like the low
$ T / |W| $ instability in a collapsing stellar core
\citep{centrella_01_a}, proto-neutron star
\citep[PNS;][]{shibata_05_b, ott_05_a, ott_07_a}, or cold neutron star
\citep[NS;]{shibata_02_a, shibata_03_a, ou_04_a, ou_06_a}. In all of
these cases, as compact objects are involved, general relativistic
(GR) effects should be taken into account properly. Moreover,
extremely high radial and/or rotational velocities may be encountered,
and thus a relativistic description of self-gravity \emph{and} of the
flow dynamics may be required. However, as an elaborate treatment of
the microphysics in the form of neutrino transport and a microphysical
equation of state (EoS) is crucial for some of the scenarios, and as
multi-dimensional GR codes for numerical simulations may either be
unavailable (as in the case of detailed Boltzmann neutrino transport)
or computationally very expensive, most previous investigations relied
on the Newtonian approximation for the strong gravitational field of
the compact objects, and on Newtonian flow dynamics.

A straightforward and easy to implement way of overcoming the first,
most serious deficiency has been suggested recently for supernova core
collapse involving at most moderately rapid rotation
\citep{rampp_02_a, marek_06_a}. The strategy presented in these studies
is to treat relativistic gravity by using an effective relativistic
gravitational potential, while retaining the equations of Newtonian
hydrodynamics. This allows for a simple upgrade of existing Newtonian
hydrodynamics codes to treat relativistic gravity in an approximate
but quite accurate manner. Extensive tests of this approach have been
presented for simulations of supernova core collapse in spherical
symmetry including Boltzmann neutrino transport and a microphysical
EoS \citep{liebendoerfer_05_a, marek_06_a}. The results of these tests
show an excellent agreement with those obtained by corresponding fully
relativistic 1D radiation hydrodynamic simulations. In addition,
\citet{marek_06_a} (Paper~I hereafter) who studied rotational core
collapse with a simplified EoS find good agreement between the results
obtained with an axisymmetric Newtonian hydrodynamics code extended by
an effective relativistic gravitational potential and those obtained
with a GR code for cores with slow and moderate rotation. However, for
rapidly rotating cores the effective relativistic potentials
considered by \citet{marek_06_a} cannot reproduce the dynamics of the
collapse correctly, and e.g.\ underestimate the maximum density at
bounce by more than $ 30\% $. In particular, their potential~A yields
excellent results in the spherical limit, but clearly fails for rapid
rotation, while potential~R performs slightly better in the latter
case, but persistently overestimates the maximum density during the
evolution for slow or no rotation.

The results of Paper~I suggest that the concept of an effective
relativistic gravitational potential can in principle be used to model
compact stars with rapid rotation. However, additional corrections to
account for stronger centrifugal forces are necessary. To this end,
we introduce here a new effective relativistic potential which
contains correction terms for the case of rapid rotation. We assess
the applicability of the new potential to the scenarios listed above,
by presenting comparisons with both GR simulations and Newtonian
simulations using two of the effective relativistic potentials
described in Paper~I for a range of core collapse and NS test
models. We explore the new effective relativistic potential in the
context of stellar core collapse with non-negligible rotation using
either a simplified EoS or a microphysical equation in combination
with an approximate deleptonization scheme. Furthermore, in order to
test whether the new effective relativistic potential is also
appropriate for compact \emph{equilibrium} configurations, we consider
rotating equilibrium polytropes which serve as models for a cold
NS. We investigate pulsations of these models and briefly discuss the
applicability of effective relativistic potentials to NS
asteroseismology.

All of the above mentioned astrophysical events are interesting
sources of gravitational radiation. Therefore, the extraction of the
gravitational wave (GW) signal from numerical simulations (possibly
using effective relativistic potentials) and the prediction of wave
templates for current and future GW detectors of interferometric and
resonant type is an important issue. To this end, we discuss the
results of \citet{obergaulinger_06_a}, who used one of the effective
relativistic potentials from Paper~I and report an overestimation of
the GW amplitudes of up to $ 50\% $ compared to simulations in GR even
for slow rotators. We resolve this discrepancy by introducing a
formulation of the Newtonian quadrupole formula that is better suited
for both Newtonian simulations with an effective relativistic
potential and fully relativistic simulations.

Throughout the article we use geometrized units with $ c = G = 1 $.


\section{Effective relativistic potential}
\label{sec:effective_potential}

In previous work \citep[][and Paper~I]{rampp_02_a} it was proposed to
approximate the effects of GR in a Newtonian hydrodynamics code by
solely substituting the Newtonian potential by an effective
relativistic gravitational potential $ \Phi_\mathrm{eff} $ which
mimics the deeper potential well of relativistic gravity. We first
recapitulate some properties of the Newtonian potential in
Sect.~\ref{subsec:potential_case_n}, of the effective relativistic
(TOV) potential introduced by \citet{rampp_02_a} in
Sect.~\ref{subsec:potential_case_r}, and of its most accurate
improvement for slow and moderate rotation from Paper~I in
Sect.~\ref{subsec:potential_case_a}. Then we present in
Sect.~\ref{subsec:potential_case_ar} the new effective relativistic
potential including rotational corrections for rapidly rotating
configurations. Finally, in Sect.~\ref{subsec:potential_case_gr} we
summarize the treatment of GR in the conformally flat approximation,
which is applied both in some previous studies and also in the present
investigation of effective relativistic potentials for supernova core
collapse for the purpose of comparison.


\subsection{Newtonian gravitational potential for a self-gravitating
  fluid}
\label{subsec:potential_case_n}

In a genuine Newtonian simulation, which we refer to as \emph{Case~N}
in the following (see also Paper~I), the Newtonian hydrodynamic
equations are solved in conjunction with the (multi-dimensional)
Newtonian gravitational potential
\begin{equation}
  \Phi_\mathrm{N} = - \int_V \!\! \mathrm{d}r' \, \mathrm{d}\theta' \,
  \mathrm{d}\varphi' \, r'^2 \sin \theta'
  \frac{\rho}{|\bm{r} - \bm{r'}|},
  \label{eq:potential_case_n}
\end{equation}
where $ \rho $ is the (rest-mass) density of the self-gravitating
fluid. This potential is the solution of the Poisson equation
\begin{equation}
  \Delta \Phi_\mathrm{N} = 4 \pi \rho.
  \label{eq:poisson_equation_case_n}
\end{equation}

For later use we also define the \emph{spherical part} of the Newtonian
potential by
\begin{equation}
  \phicapbar_\mathrm{N} (r) =
  - \int_r^\infty \frac{\mathrm{d}r'}{r'^2} \overline{m}_\mathrm{N}
  \label{eq:potential_spherical_case_n}
\end{equation}
and the spherical Newtonian mass as
\begin{equation}
  \overline{m}_\mathrm{N} (r) = 4 \pi \int_0^r \mathrm{d}r' \, r'^2
  \left\langle \rho \right\rangle,
  \label{eq:mass_spherical_case_n}
\end{equation}
where any hydrodynamic quantity enclosed by brackets,
$ \langle \dots \rangle $, is the angular average of its
multi-dimensional counterpart.


\subsection{TOV potential for a self-gravitating fluid}
\label{subsec:potential_case_r}

For devising an effective relativistic potential for a
self-gravitating perfect fluid we require that for a spherically
symmetric configuration it should reproduce the
Tolman--Oppenheimer--Volkoff (TOV) solution of hydrostatic equilibrium
in GR. Then a first approximation to include relativistic effects in a
Newtonian code is to replace the spherical part $ \phicapbar (r) $ of
the gravitational potential $ \Phi (r, \theta, \phi) $ by the
\emph{spherical TOV potential} $ \phicapbar_\mathrm{TOV} $
\citep{rampp_02_a}:
\begin{equation}
  \Phi_\mathrm{R} = \Phi_\mathrm{N} - \phicapbar_\mathrm{N} +
  \phicapbar_\mathrm{TOV}.
  \label{eq:potential_case_r}
\end{equation}
As in Paper~I we refer to this effective relativistic potential as
\emph{Case~R} (or ``reference'' case). The spherical TOV potential is
defined as
\begin{equation}
  \phicapbar_\mathrm{TOV} (r) =
   - \int_r^\infty \frac{\mathrm{d}r'}{r'^2}
  \left( \overline{m}_\mathrm{TOV} +
  4 \pi r'^3 \left\langle P \right\rangle \right)
  \frac{\left\langle h \right\rangle}{\overline{\Gamma}_\mathrm{TOV}^2},
  \label{eq:potential_spherical_case_r}
\end{equation}
where $ h = 1 + \epsilon + P / \rho $ is the relativistic specific
enthalpy, $ \epsilon $ is the specific internal energy, and $ P $ is
the fluid pressure. The spherical TOV mass is given by
\begin{equation}
  \overline{m}_\mathrm{TOV} (r) = 4 \pi \int_0^r \mathrm{d}r' \, r'^2
  \left\langle \rho \right\rangle
  \left( 1 + \left\langle \epsilon \right\rangle \right),
  \label{eq:mass_spherical_case_r}
\end{equation}
and
\begin{equation}
  \overline{\Gamma}_\mathrm{TOV} (r) =
  \sqrt{1 + \left\langle v_r \right\rangle^2 -
  \frac{2 \overline{m}_\mathrm{TOV}}{r}},
  \label{eq:gamma_spherical_case_r}
\end{equation}
with $ v_r $ being the Eulerian radial fluid velocity.


\subsection{Improved potential for a self-gravitating fluid without
  rotational corrections}
\label{subsec:potential_case_a}

\citet{marek_06_a} found in their study that the effective relativistic
potential $ \Phi_\mathrm{R} $ overestimates the GR effects due to the
straightforward combination of relativistic gravity and Newtonian
dynamics. Thus, they proposed another effective relativistic potential
\begin{equation}
  \Phi_\mathrm{A} \equiv
  \Phi_\mathrm{N} - \phicapbar_\mathrm{N} + \phicapbar_\mathrm{TOV\!,A},
  \label{eq:potential_case_a}
\end{equation}
labeled as \emph{Case~A}, which involves the \emph{modified spherical
  TOV potential}
\begin{equation}
  \phicapbar_\mathrm{TOV\!,A} (r) =
  - \int_r^\infty \frac{\mathrm{d}r'}{r'^2}
  \left( \overline{m}_\mathrm{TOV\!,A} +
  4 \pi r'^3 \left\langle P \right\rangle \right)
  \frac{\left\langle h \right\rangle}
  {\overline{\Gamma}_\mathrm{TOV\!,A}^2}.
  \label{eq:potential_spherical_case_a}
\end{equation}
The latter differs from $ \phicapbar_\mathrm{TOV} (r) $ in
Eq.~(\ref{eq:potential_spherical_case_r}) by the modified spherical
TOV mass
\begin{equation}
  \overline{m}_\mathrm{TOV\!,A} (r) = 
  4 \pi \int_0^r \mathrm{d}r' \, r'^2
  \left\langle \rho \right\rangle
  (1 + \left\langle \epsilon \right\rangle) \,
  \overline{\Gamma}_\mathrm{TOV\!,A},
  \label{eq:mass_spherical_case_a}
\end{equation}
which compared to $ \overline{m}_\mathrm{TOV} (r) $ in
Eq.~(\ref{eq:mass_spherical_case_r}) contains an additional factor
\begin{equation}
  \overline{\Gamma}_\mathrm{TOV\!,A} (r) = 
  \sqrt{1 + \left\langle v_r \right\rangle^2 -
  \frac{2 \overline{m}_\mathrm{TOV\!,A}}{r}}.
  \label{eq:gamma_spherical_case_a}
\end{equation}
Note that this factor also appears explicitly in the definition of
the modified spherical TOV potential
$ \phicapbar_\mathrm{TOV\!,A} (r) $ in
Eq.~(\ref{eq:potential_spherical_case_a}).

Since $ \overline{\Gamma}_\mathrm{TOV\!,A} < 1 $ the modified
spherical TOV mass $ \overline{m}_\mathrm{TOV\!,A} $ is smaller than
$ \overline{m}_\mathrm{TOV} $. The fact that the right hand side of
Eq.~(\ref{eq:mass_spherical_case_a}), which defines the modified
spherical TOV mass $ \overline{m}_\mathrm{TOV\!, A} $, depends itself
on $ \overline{m}_\mathrm{TOV\!,A} $, because of the factor
$ \overline{\Gamma}_\mathrm{TOV\!,A} $, causes no problem, as
$ \overline{m}_\mathrm{TOV\!,A} $ can be computed in a straightforward
way by means of an ODE integration.

As demonstrated in Paper~I, for simulations of rotational stellar core
collapse the usage of the effective relativistic potential
$ \Phi\,_\mathrm{A} $ results in collapse dynamics and a PNS structure
that are very close to the outcome in GR, at least in the regime of
slow and moderately fast rotation.


\subsection{Effective relativistic potential with rotational
  corrections}
\label{subsec:potential_case_ar}

Following the concept of effective relativistic potentials discussed
in Paper~I, which are both easy to calculate and to implement in an
existing Newtonian code, we require two criteria to be fulfilled for
an effective relativistic potential $ \Phi_\mathrm{Arot} $ including
corrections due to rotation (referred to as \emph{Case~Arot} in the
following). Firstly, any modification should only involve the changes
of the gravitational potential, thereby guaranteeing that the
structure of the term $ - \rho \nabla \Phi_\mathrm{A} $ in the
momentum equation remains unchanged and is still integrable, and also
that the Newtonian hydrodynamic equations remain unaltered. Secondly,
in order to retain the proven accuracy of the effective relativistic
potential $ \Phi_\mathrm{A} $, the new potential
$ \Phi_\mathrm{Arot} $ should reduce to that potential in the
spherical limit.

For Cases~R and~A the effective relativistic potentials are computed
by replacing only the monopole $ \phicapbar_\mathrm{N} $ of the
Newtonian gravitational field $ \Phi_\mathrm{N} $ with
$ \phicapbar_\mathrm{TOV} $ or $ \phicapbar_\mathrm{TOV\!,A} $,
respectively. The obvious restriction of this approach is that any
non-spherical (and in particular rotational) effects are accounted for
exclusively in the higher multipoles of the Newtonian potential
$ \Phi_\mathrm{N} $, limiting the GR corrections to the monopole. In
contrast, Case~Arot also involves GR corrections to the multipoles by
modifying the source term of the Poisson
equation~(\ref{eq:poisson_equation_case_n}).

Hence, we introduce a \emph{modified Newtonian potential}
$ \Phi_\mathrm{N,Arot} $, whose spherical part
$ \phicapbar_\mathrm{N,Arot} $ is (as in the other cases) substituted
by the corresponding modified spherical TOV potential
$ \phicapbar_\mathrm{TOV\!,Arot} $:
\begin{equation}
  \Phi_\mathrm{Arot} \equiv \Phi_\mathrm{N,Arot} -
  \phicapbar_\mathrm{N,Arot} +
  \phicapbar_\mathrm{TOV\!,Arot}.
  \label{eq:potential_case_ar}
\end{equation}
The modified Newtonian potential $ \Phi_\mathrm{N,Arot} $ is
calculated by applying several changes to the source term in the
Poisson equation~(\ref{eq:poisson_equation_case_n}). As a first step,
we replace the density $ \rho $ by $ \rho W_\Omega $, where the
angular Lorentz factor $ W_\Omega = (1 - v_\Omega^2)^{-1/2} $ with
$ v_\Omega^2 = v_\theta^2 + v_\varphi^2 $ is restricted to the
meridional and rotational velocity components $ v_\theta $ and
$ v_\varphi $. With this choice, in the case of a purely angular
velocity the density equals the conserved density in the GR continuity
equation in the ADM $ 3 + 1 $ split \citep[see
e.g.][]{banyuls_97_a}. Disregarding the radial velocity component
$ v_r $ in the Lorentz factor is motivated by the Birkhoff theorem,
according to which neither the gravitational mass nor the vacuum field
of a spherical star change if the matter is subject to radial motion
only.

A comparison of the spherical Newtonian potential
$ \phicapbar_\mathrm{N} $ with the (unmodified) spherical TOV
potential $ \phicapbar_\mathrm{TOV} $ suggests additional GR
corrections for the source term of $ \Phi_\mathrm{N,Arot} $.
Neglecting the pressure term in
Eq.~(\ref{eq:potential_spherical_case_r}) we get
\begin{equation}
  r^2 \frac{\partial \phicapbar_\mathrm{TOV}}{\partial r} =
  \overline{m}_\mathrm{TOV} \,
  \frac{\left\langle h \right\rangle}{\overline{\Gamma}_\mathrm{TOV}^2},
  \label{eq:potential_spherical_case_r_differential_form}
\end{equation}
and from Eq.~(\ref{eq:potential_spherical_case_n})
\begin{equation}
  r^2 \frac{\partial \phicapbar_\mathrm{N}}{\partial r} = 
  \overline{m}_\mathrm{N}.
  \label{eq:potential_spherical_case_n_differential_form}
\end{equation}
By identifying $ \overline{m}_\mathrm{TOV} $ as a relativistic
generalization of $ \overline{m}_\mathrm{N} $, we obtain an additional
correction factor $ h / \Gamma_\mathrm{TOV\!,Arot}^2 $, which
approximately takes into account the influence of the relativistic
enthalpy $ h $ and the self-interaction of the gravitational field in
GR through $ \Gamma_\mathrm{TOV\!,Arot} $. The factor $ h $ can also
be inferred by comparing the gravity terms $ \nabla \Phi_\mathrm{N} $
and $ h \nabla \nu $ in the equations of rotating equilibrium
configurations in Newtonian and GR gravity, respectively, where
$ \nu $ is a component of the spacetime metric which corresponds to
the gravitational potential $ \Phi_\mathrm{N} $ in Newtonian gravity.

Thus, for calculating $ \Phi_\mathrm{N,Arot} $ we finally arrive at
\begin{equation}
  \Delta \Phi_\mathrm{N,Arot} = 4 \pi \rho_\mathrm{Arot},
  \label{eq:poisson_equation_case_ar_n}
\end{equation}
with the following expression for the source term:
\begin{equation}
  \rho_\mathrm{Arot} =
  \frac{\rho h W_\Omega}{\Gamma_\mathrm{TOV\!,Arot}^2}.
  \label{eq:density_case_ar_n}
\end{equation}
Accordingly, the effective relativistic potential
$ \Phi_\mathrm{N,Arot} $ can then be calculated in integral form
as
\begin{equation}
  \Phi_\mathrm{N,Arot} =
  - \int_V \!\! \mathrm{d}r' \, \mathrm{d}\theta' \,
  \mathrm{d}\varphi' \, r'^2 \sin \theta'
  \frac{\rho_\mathrm{Arot}}{|\bm{r} - \bm{r'}|}
  \label{eq:potential_case_ar_n}
\end{equation}
and its spherical part as
\begin{equation}
  \phicapbar_\mathrm{N,Arot} (r) =
  - \int_0^\infty \frac{\mathrm{d}r'}{r'^2} \overline{m}_\mathrm{N,Arot}.
  \label{eq:potential_spherical_case_ar_n}
\end{equation}
Here the modified spherical Newtonian mass is defined as
\begin{equation}
  \overline{m}_\mathrm{N,Arot} (r) =
  4 \pi \int_0^r \mathrm{d}r' \, r'^2 \, \rhobar_\mathrm{Arot},
  \label{eq:mass_spherical_case_ar_n}
\end{equation}
with the modified spherical density $ \rhobar_\mathrm{Arot} $ given
by
\begin{equation}
  \rhobar_\mathrm{Arot} =
  \frac{\left\langle \rho \right\rangle \!
  \left\langle h \right \rangle \!
  \left\langle W_\Omega \right\rangle \!} 
  {\overline{\Gamma}_\mathrm{TOV\!,Arot}^2}.
  \label{eq:density_spherical_case_ar_n}
\end{equation}
Note that $ \rhobar_\mathrm{Arot} $ is not computed as the angular
average of $ \rho_\mathrm{Arot} $ itself but rather of its
constituents, because it is a quantity derived from other hydrodynamic
quantities. The angular average of the Lorentz factor is given by
$ \langle W_\Omega \rangle = \langle (1 - v_\Omega^2)^{-1/2} \rangle $.

For the modified spherical TOV potential of Case~Arot we use an
expression that is analogous to
Eqs.~(\ref{eq:potential_spherical_case_r}) and
(\ref{eq:potential_spherical_case_a}):
\begin{equation}
  \begin{array}{rcl}
    \displaystyle
    \phicapbar_\mathrm{TOV\!,Arot} (r) & = &
    \displaystyle
    - \int_r^\infty \frac{\mathrm{d}r'}{r'^2}
    \left( \overline{m}_\mathrm{TOV,Arot} +
    4 \pi r'^3 \left\langle P \right\rangle \right) \cdot
    \\ [0.8 em]
    & & \displaystyle \qquad \quad
    \frac{\left\langle h \right\rangle}{\overline{\Gamma}_\mathrm{TOV\!,Arot}^2}
    \left\langle\! (1 - v_\Omega^2)^{-1} \!\right\rangle_{\!\!\rho}.
  \end{array}
  \label{eq:potential_spherical_case_ar_r}
\end{equation}
This potential also contains several additions related to
non-spherical effects. First of all, in the modified spherical TOV
mass
\begin{equation}
  \overline{m}_\mathrm{Arot} (r) =
  4 \pi \int_0^r \mathrm{d}r' \, r'^2
  \left\langle \rho \:\! W_\Omega \right\rangle
  (1 + \left\langle \epsilon \right\rangle)
  \overline{\Gamma}_\mathrm{Arot},
  \label{eq:mass_spherical_case_ar}
\end{equation}
the angular averaged density $ \langle \rho \rangle $ is
substituted by $ \langle \rho W_\Omega \rangle $ with the same
rationale as in the source term of the Poisson equation
(\ref{eq:poisson_equation_case_ar_n}), while the definition of
\begin{equation}
  \overline{\Gamma}_\mathrm{TOV,Arot} (r) =
  \sqrt{1 + \left\langle v_r \right\rangle^2 -
  \frac{2 \overline{m}_\mathrm{TOV,Arot}}{r}}
  \label{eq:gamma_spherical_case_ar}
\end{equation}
follows those of Eqs.~(\ref{eq:gamma_spherical_case_r}) and
(\ref{eq:gamma_spherical_case_a}). The introduction of the
centrifugal factor
\begin{equation}
  \left\langle \! (1 - v_\Omega^2)^{-1} \! \right\rangle_{\!\!\rho} =
  \frac{\left\langle \rho W^3_\Omega \right\rangle}
  {\left\langle \rho W_\Omega \right\rangle}
\end{equation}
in the effective relativistic
potential~(\ref{eq:potential_spherical_case_ar_r}) is
suggested by comparing the equations for stationary equilibria of
rotating self-gravitating fluids in Newtonian and GR gravity. In
essence, the centrifugal factor results in a larger gravitational
acceleration for high rotational velocities. It mimics a relativistic
effect that is, for instance, known from the motion of a test particle
in a Schwarzschild background spacetime, where the effective
relativistic potential \citep[see e.g.][]{straumann_04_a}
\begin{equation}
  V_\mathrm{eff} = \left( 1 - \frac{2 m}{r} \right)
  \left( 1 + \frac{l^2}{r^2} \right)
\end{equation}
for the motion of a test particle with specific angular momentum
$ l = r v_\Omega $ contains a factor $ (1 + l^2 / r^2) $, which
is identical to $ (1 - v_\Omega^2)^{-1} $ to order
$ \mathcal{O} (v^2) $. The density averaging is preferable to a volume
averaging to account for the rotational flattening in cases of rapid
rotation.

As required the new effective relativistic potential
$ \Phi_\mathrm{Arot} $ by construction reduces to $ \Phi_\mathrm{A} $
in spherical symmetry. In this limit the multi-dimensional and
spherical (modified) Newtonian potentials are identical,
$ \Phi_\mathrm{N} = \phicapbar_\mathrm{N} $ and
$ \Phi_\mathrm{N,Arot} = \phicapbar_\mathrm{N,Arot} $. Thus, they
cancel each other in Eqs.~(\ref{eq:potential_case_a})
and~(\ref{eq:potential_case_ar}), respectively, while the modified TOV
potentials~(\ref{eq:potential_spherical_case_a},
\ref{eq:potential_spherical_case_ar_r}) become identical as
$ \langle W_\Omega \rangle $ vanishes.

Note that different from Paper~I, for reason of simplicity we omit all
contributions due to neutrino effects in the effective relativistic
potentials discussed here. However, it is straightforward to add the
appropriate terms to any of the potentials (see Paper~I).


\subsection{General relativistic formulation using the conformally
  flat approximation}
\label{subsec:potential_case_gr}

To be able to directly assess the quality of the effective
relativistic potential approach in describing rapidly rotating compact
objects, we have repeated all simulations using a GR hydrodynamics
code. These results are referred to as \emph{Case~GR}. As in Paper~I,
we approximate the GR spacetime metric by the conformal flatness
condition \citep[CFC; see][]{isenberg_78_a, wilson_96_a}, whose
excellent quality in the context of rotational stellar core collapse
and a rotating NS has been demonstrated extensively \citep[see e.g.]
[and references therein]{shibata_04_a, cerda_05_a, ott_07_a}.


\section{Numerical methods and models}
\label{sec:methods}

In order to comprehensively test the new effective relativistic
potential in situations where centrifugal forces play an important
role, we apply it to simulations of the collapse of a rotating stellar
core to a PNS and of a rapidly rotating NS in equilibrium. In
Sect.~\ref{subsec:code} we summarize the properties of the code used
for these simulations, while in Sect.~\ref{subsec:models} we describe
the setup of the initial models and the EoS for each of the two
scenarios.


\subsection{Code and grid setup}
\label{subsec:code}

All simulations are performed with the code \textsc{CoCoNuT} of
\citet{dimmelmeier_02_a, dimmelmeier_05_a} assuming axial symmetry and
symmetry with respect to the equatorial plane. This code optionally
either uses a Newtonian (or alternatively an effective relativistic)
gravitational potential and Newtonian hydrodynamics (Cases~N, R, A,
and~Arot) \emph{or} solves the GR equations of fluid dynamics and the
GR field equations in the ADM $ 3 + 1 $ split assuming CFC (Case~GR).
This procedure allows for a direct comparison of the different
effective relativistic potentials with GR using one numerical code and
an identical grid setup.

The \textsc{CoCoNuT} code employs a metric solver based on spectral
methods as described in \citet{dimmelmeier_05_a}. The (Newtonian or
GR) hydrodynamic equations are formulated in conservation form, and
are solved by high-resolution shock-capturing schemes with
state-of-the-art Riemann solvers and piecewise parabolic cell
reconstruction procedures on an Eulerian grid in spherical polar
coordinates $ \{r, \theta \} $.

Rotating NS are modelled with an equidistant grid of $ 80 \times 30 $
zones. For the simulations of stellar core collapse the radial grid is
logarithmically spaced to increase the effective resolution in the
center where the PNS forms. In that case the grid consists of
$ 200 \times 30 $ zones in $ \{r, \theta\} $ with a central radial
resolution of $ \Delta r_\mathrm{c} = 500 \mathrm{\ m} $ for models
with a simplified EoS, and of $ 250 \times 45 $ zones with
$ \Delta r_\mathrm{c} = 250 \mathrm{\ m} $ for the models computed
with a microphysical EoS. 
In both cases, the computational domain extends
from $r_\mathrm{min}=0$ to $r_\mathrm{max}=2000\mathrm{\ km}$. Reflecting
boundary conditions are imposed at the origin; otherwise no specific
measures are taken to deal with the singularity at $r=0$. A small part of the
grid is occupied by an artificial low-density atmosphere extending beyond
the surface of the NS or stellar core.
For more details about the
grid setup and the boundary conditions, we refer to e.g.\
\citet{dimmelmeier_05_a}, \citet{dimmelmeier_06_a}, or \citet{ott_07_b}.


\subsection{Equations of state, and models for rotating neutron stars
  or stellar core collapse}
\label{subsec:models}

\begin{table*}[t]
{
  \caption{Nomenclature of the iron core collapse models AxByGz
    (simulated with the hybrid EoS) and s20AxBy (simulated with the
    microphysical EoS). The length parameter $ A $ specifies the
    initial degree of differential rotation, $ (T / |W|)_\mathrm{i} $ is
    the initial rotation rate, and $ \gamma_1 $ is the stiffness of
    the subnuclear part of the hybrid EoS, respectively. We also
    give the angular velocity at the center for the initial configuration
    ($\Omega_\mathrm{c,i}$) and at bounce ($\Omega_\mathrm{c,b}$).
    For models marked with an asterisk, we give the maximum angular
    velocity $\Omega_\mathrm{max,b}$ instead of $\Omega_\mathrm{c,b}$,
    because the density in the central region is extremely low due
    to the toroidal density stratification, and 
    $\Omega_\mathrm{c,b}$ is therefore no good indicator for the
    rotational state of the entire core.}
  \label{tab:models}
}
  \begin{center}
    \begin{tabular}{lccccc}
      \hline \hline
      Model &
      $ A \mathrm{\ [1000 km]} $ &
      $ (T / |W|)_\mathrm{i} \mathrm{\ [\%]} $ &
      $ \Omega_\mathrm{c,i} \mathrm{\ [s^{-1}]}$ &
      $ \Omega_\mathrm{c,b} \mathrm{\ [10^{3} s^{-1}]}$ &
      $ \gamma_1 $ \\
      \hline
      A1B1G1  &  50.0 & 0.25  & \z\z2.3 & $ 2.9\a $ & 1.325 \\
      A1B2G1  &  50.0 & 0.5\z & \z\z3.1 & $ 3.6\a $ & 1.325 \\
      A1B3G1  &  50.0 & 0.9\z & \z\z4.1 & $ 6.2\a $ & 1.325 \\
      A1B3G2  &  50.0 & 0.9\z & \z\z4.1 & $ 4.8\a $ & 1.320 \\
      A1B3G3  &  50.0 & 0.9\z & \z\z4.1 & $ 2.8\a $ & 1.310 \\ [0.5 em]
      A2B4G1  & \z1.0 & 1.8\z & \z\z8.0 & $ 1.8\a $ & 1.325 \\ [0.5 em]
      A3B1G1  & \z0.5 & 0.25  & \z\z4.6 & $ 3.9\a $ & 1.325 \\
      A3B2G4  & \z0.5 & 0.5\z & \z\z6.3 & $ 6.6\a $ & 1.300 \\
      A3B4G2  & \z0.5 & 1.8\z & \z 12.4 & $ 2.3\a $ & 1.320 \\ [0.5 em]
      A4B1G1  & \z0.1 & 0.25  & \z 31.1 & $ 21 \a $ & 1.325 \\
      A4B1G2  & \z0.1 & 0.25  & \z 31.1 & $ 21 \a $ & 1.320 \\
      A4B2G2  & \z0.1 & 0.5\z & \z 43.6 & $ 22 \a $ & 1.320 \\
      A4B2G3  & \z0.1 & 0.5\z & \z 43.6 & $ 7.5\a $ & 1.310 \\
      A4B5G4  & \z0.1 & 4.0\z &   139.4 & $ 2.4^* $ & 1.300 \\
      A4B5G5  & \z0.1 & 4.0\z &   139.4 & $ 2.5^* $ & 1.280 \\ [0.5 em]
      s20A1B1 &  50.0 & 0.25  & \z\z1.0 & $ 1.0\a $ & ---   \\
      s20A2B4 & \z1.0 & 0.5\z & \z\z9.6 & $ 3.3\a $ & ---   \\
      \hline \hline
    \end{tabular}
  \end{center}
\end{table*}

As initial data for the simulations of NS we choose rotating
equilibrium configurations which obey a polytropic EoS,
\begin{equation}
  P = K \rho^\gamma
  \label{eq:polytrope}
\end{equation}
with $ K = 1.4553 \times 10^5 $ (in cgs units\footnote{This value is
  commonly used for polytropic NS models.}) and $ \gamma = 2 $. The
models are computed with the self-consistent field method of
\citet{komatsu_89_a}. Their rotation law for the specific angular
momentum $ j $ is given by
\begin{equation}
  j = A^2 (\Omega_\mathrm{c} - \Omega),
  \label{eq:rotation_law}
\end{equation}
where $ A $ parameterizes the degree of differential rotation (more
differential for smaller values of $ A $), and $ \Omega_\mathrm{c} $
is the value of the angular velocity $ \Omega $ at the center. In the
Newtonian limit, this rotation law reduces to
\begin{equation}
  \Omega = \frac{A^2 \Omega_\mathrm{c}}{A^2 + \varpi^2},
  \label{neq:ewtonian_rotation_law}
\end{equation}
where $ \varpi $ is the distance to the rotation axis.

In the simulations of rotational core collapse with a simplified EoS
the initial rotating equilibrium models are computed with a polytropic
EoS. They approximate an iron core supported by electron degeneracy
pressure with a central density
$ \rho_\mathrm{c,i} = 10^{10} \mathrm{\ g\ cm}^{-3} $, and the EoS
parameters $ K = 4.897 \times 10^{14} $ and $ \gamma = 4 / 3 $,
respectively. For the evolution of the cores a hybrid EoS is employed
\citep{janka_93_a, dimmelmeier_02_a}. It consists of a polytropic
contribution describing the degenerate electron pressure and (at
supranuclear densities) the pressure due to repulsive nuclear forces,
and a thermal contribution which accounts for shock heating:
\begin{equation}
  P = P_\mathrm{p} + P_\mathrm{th},
  \label{eq:hybrid_eos}
\end{equation}
where
\begin{equation}
  P_\mathrm{p} = K \rho^{\gamma},
  \qquad
  P_\mathrm{th} = \rho \epsilon_\mathrm{th} (\gamma_\mathrm{th} - 1),
  \label{eq:hybrid_eos_terms}
\end{equation}
with $ \gamma_\mathrm{th} = 1.5 $. To trigger the collapse the
adiabatic index is reduced from its initial value $ 4 / 3 $ to
$ \gamma_1 < 4 / 3 $. At nuclear density
$ \rho_\mathrm{nuc} = 2 \times 10^{14} \mathrm{\ g\ cm}^{-3} $ the
adiabatic index is raised abruptly to a value $ \gamma_2 = 2.5 $ to
mimic the stiffening of the nuclear EoS, which results in a rebound of
the core (the core bounce) and the formation of the PNS. The hybrid
EoS provides for a smooth transition between these two density regimes
\citep[for more details, see][]{janka_93_a}.

In our microphysically more detailed calculations of stellar core
collapse we employ the tabulated non-zero temperature nuclear EoS by
\citet{shen_98_a} in the variant of \citet{marek_05_a} which includes
pressure contributions from baryons, leptons, and photons.
Deleptonization by electron capture onto nuclei and free protons is
implemented according to \citet{liebendoerfer_05_b}: During collapse
the electron fraction $ Y_e $ is parameterized as a function of
density based on neutrino radiation-hydrodynamic simulations in
spherical symmetry \citep{marek_05_a} using the electron capture rates
of \citep{langanke_00_a}. After core bounce, $ Y_e $ is only passively
advected, and any further lepton loss is neglected. Again following
\citet{liebendoerfer_05_b}, above the trapping density
$ \rho_\mathrm{trap} = 2.0 \times 10^{12} \mathrm{\ g\ cm}^{-3} $
contributions due to neutrino radiation pressure are taken into
account. As initial models we take the non-rotating $ 20 \, M_\odot $
solar-metallicity progenitor s20.0 from \citet{woosley_02_a} and
impose the rotation law~(\ref{eq:rotation_law}). Note that these
initial models are not in equilibrium, because the progenitor model
s20.0 already shows considerable negative radial velocities and the
models are not relaxed with respect to the rotation added.

\begin{table*}[t]
  \caption{Maximum density $ \rho_\mathrm{max,b} $ at bounce and final
    maximum density $ \rho_\mathrm{max,f} $ after ring-down in units
    of $ 10^{14} \mathrm{\ g\ cm}^{-3} $ for all investigated core
    collapse models in Case~GR, or evolved with the effective
    relativistic potential $ \Phi_\mathrm{R} $, and
    $ \Phi_\mathrm{Arot} $, respectively. For each potential, the
    relative deviation from Case~GR (in percent) is given in
    parentheses, and the value closest to Case~GR is indicated by an
    asterisk.     Note that some centrifugally
    bouncing models do not reach stable values for
    $ \rho_\mathrm{max,f} $ until very long after core bounce.}
  \label{tab:results_density}

  \begin{center}
    \begin{tabular}{l@{\qquad}cccc@{\qquad}cccc}
      \hline \hline
      & \multicolumn{4}{c}{$ \rho_\mathrm{max,b} $} &
      \multicolumn{4}{c}{$ \rho_\mathrm{max,f} $} \\ [0.5 em]
      Model & Case~GR & Case~R & Case~A & Case~Arot &
      Case~GR & Case~R & Case~A & Case~Arot \\
      \hline
      A1B1G1   &
      $ 4.71 $ &
      $ 4.55 ~ (-3)^*\z $ &
      $ 4.32 ~ (-8)\a\z $ &
      $ 4.42 ~ (-6)\a\z $ &
      $ 3.30 $ &
      $ 3.53 ~ (+7)\a\z $ &
      $ 3.36 ~ (+2)^*\z $ &
      $ 3.40 ~ (+3)\a\z $ \\
      A1B2G1 &
      $ 4.37 $ &
      $ 4.26 ~ (-3)^*\z $ &
      $ 3.98 ~ (-9)\a\z $ &
      $ 4.16 ~ (-5)\a\z $ &
      $ 3.16 $ &
      $ 3.32 ~ (+5)\a\z $ &
      $ 3.19 ~ (+1)^*\z $ &
      $ 3.25 ~ (+3)\a\z $ \\
      A1B3G1 &
      $ 3.86 $ &
      $ 3.78 ~ (-2)\a\z $ &
      $ 3.56 ~ (-7)\a\z $ &
      $ 3.81 ~ (-1)^*\z $ &
      $ 2.84 $ &
      $ 3.00 ~ (+6)\a\z $ &
      $ 2.82 ~ (-1)^*\z $ &
      $ 2.97 ~ (+5)\a\z $ \\
      A1B3G2 &
      $ 4.02 $ &
      $ 3.88 ~ (-3)\a\z $ &
      $ 3.70 ~ (-8)\a\z $ &
      $ 3.90 ~ (-3)^*\z $ &
      $ 2.78 $ &
      $ 2.88 ~ (+4)\a\z    $ &
      $ 2.78 ~ (\pm 0)^*\z $ &
      $ 2.84 ~ (+2)\a\z    $ \\
      A1B3G3 &
      $ 4.19 $ &
      $ 4.04 ~ (-4)^*\z $ &
      $ 3.93 ~ (-6)\a\z $ &
      $ 4.01 ~ (-4)\a\z $ &
      $ 2.99 $ &
      $ 2.11 ~ (+4)\a\z $ &
      $ 3.03 ~ (+1)^*\z $ &
      $ 3.10 ~ (+4)\a\z $ \\ [0.5 em]
      A2B4G1 &
      $ 0.64 $ &
      $ 0.44 ~ (-31)\a  $ &
      $ 0.30 ~ (-53)\a  $ &
      $ 0.58 ~ (-9)^*\z $ &
      --- &
      --- &
      --- &
      --- \\ [0.5 em]
      A3B1G1 &
      $ 4.46 $ &
      $ 4.27 ~ (-4)^*\z $ &
      $ 4.05 ~ (-9)\a\z $ &
      $ 4.25 ~ (-5)\a\z $ &
      $ 3.16 $ &
      $ 3.33 ~ (+5)\a\z $ &
      $ 3.18 ~ (+1)^*\z $ &
      $ 3.26 ~ (+4)\a\z $ \\
      A3B2G4 &
      $ 3.98 $ &
      $ 3.83 ~ (-4)\a\z $ &
      $ 3.76 ~ (-6)\a\z $ &
      $ 3.87 ~ (-3)^*\z $ &
      $ 2.54 $ &
      $ 2.57 ~ (+1)\a\z    $ &
      $ 2.54 ~ (\pm 0)^*\z $ &
      $ 2.60 ~ (+2)\a\z    $ \\
      A3B4G2 &
      $ 0.59 $ &
      $ 0.41 ~ (-31)\a  $ &
      $ 0.33 ~ (-44)\a  $ &
      $ 0.58 ~ (-1)^*\z $ &
      --- &
      --- &
      --- &
      --- \\ [0.5 em]
      A4B1G1 &
      $ 4.64 $ &
      $ 4.16 ~ (-7)\a\z $ &
      $ 3.94 ~ (-15)\a  $ &
      $ 4.30 ~ (-10)^*  $ &
      $ 2.98 $ &
      $ 2.96 ~ (-1)^*\z $ &
      $ 2.85 ~ (+4)\a\z $ &
      $ 3.09 ~ (+4)\a\z $ \\
      A4B1G2 &
      $ 4.46 $ &
      $ 4.12 ~ (-8)\a\z $ &
      $ 3.88 ~ (-7)\a\z $ &
      $ 4.23 ~ (-5)^*\z $ &
      $ 2.84 $ &
      $ 2.83 ~ (\pm 0)^*\z $ &
      $ 2.75 ~ (-3)\a\z    $ &
      $ 2.96 ~ (+4)\a\z    $ \\
      A4B2G2 &
      $ 4.31 $ &
      $ 3.59 ~ (-17)\a  $ &
      $ 3.47 ~ (-20)\a  $ &
      $ 4.08 ~ (-2)^*\z $ &
      $ 2.56 $ &
      $ 2.37 ~ (-7)\a\z $ &
      $ 2.30 ~ (-10)\a  $ &
      $ 2.72 ~ (+6)^*\z $ \\
      A4B2G3 &
      $ 4.00 $ &
      $ 4.46 ~ (+11)\a  $ &
      $ 3.38 ~ (-16)\a  $ &
      $ 3.91 ~ (-5)^*\z $ &
      $ 2.30 $ &
      $ 2.14 ~ (-7)\a\z $ &
      $ 1.55 ~ (-33)\a  $ &
      $ 2.42 ~ (+5)^*\z $ \\
      A4B5G4 &
      $ 0.47 $ &
      $ 0.36 ~ (-25)\a  $ &
      $ 0.35 ~ (-27)\a  $ &
      $ 0.51 ~ (+7)^*\z $ &
      --- &
      --- &
      --- &
      --- \\
      A4B5G5 &
      $ 1.72 $ &
      $ 1.26 ~ (-27) \a    $ &
      $ 1.26 ~ (-27) \a    $ &
      $ 1.72 ~ (\pm 0)^*\z $ &
      --- &
      --- &
      --- &
      --- \\ [0.5 em]
      s20A1B1 &
      $ 3.21 $ &
      $ 3.14 ~ (-2)^*\z $ &
      $ 3.06 ~ (-5)\a\z $ &
      $ 3.07 ~ (-4)\a\z $ &
      $ 2.69 $ &
      $ 2.78 ~ (+3)\a\z $ &
      $ 2.73 ~ (+1)^*\z $ &
      $ 2.74 ~ (+2)\a\z $ \\
      s20A2B4 &
      $ 1.61 $ &
      $ 1.18 ~ (-27)\a   $ &
      $ 0.89 ~ (-45)\a   $ &
      $ 1.64 ~ (+2)^* \z $ &
      $ 0.94 $ &
      $ 0.75 ~ (-20)^* $ &
      $ 0.62 ~ (-34)\a $ &
      $ 1.18 ~ (+26)\a $ \\
      \hline \hline
    \end{tabular}
  \end{center}
\end{table*}

In Table~\ref{tab:models} we summarize the parameters of all
investigated stellar core collapse models. They are characterized by
their initial degree of differential rotation $ A $ and by their initial
rotation parameter $(T / |W|)_\mathrm{i} $, which is the
ratio of rotational energy to (the absolute value of the)
gravitational binding energy. The corresponding angular velocity
$\Omega_\mathrm{c,i}$ of the progenitor model is also shown, as
well as the angular velocity $\Omega_\mathrm{c,b}$ at bounce.
For models evolved using the hybrid EoS, the stiffness
$\gamma_1$ of the subnuclear EoS is also specified.

Our selection of models includes fast-rotating cores which end up as PNSs with
spin periods in the milisecond or sub-milisecond range. Lower initial rotation rates --
as predicted by recent stellar evolution models incorporating magnetic
braking \citep{heger_05_a} --  are required to match the spin periods of
observed young pulars, which are typically larger than $10 \mathrm{\ ms}$. 
However, since our goal is to test the effective potential approach
for such rotation rates as required by the scenarios (collapsars, etc.)
mentioned in the introduction, we also consider rapidly rotating models. While
rare in nature, there are viable formation channels for such
progenitors: Even when magnetic torques are included, it is still possible
to obtain fast rotating cores for massive Wolf--Rayet stars for sufficiently
low mass loss rates \citep{woosley_06_a} (which may be realistic
for low-metallicity stars), although few supernovae are expected
to originate from such progenitors because of the declining initial mass
function. Binary evolution effects, i.e.\ accretion
\citep{cantiello_07_a, yoon_05_a} or mergers \citep{fryer_05_a}, may
also lead to rapidly spinning cores. Again, only a small fraction of
core collapse events can be explained this way: For the particular
case of accretion-induced collapse, the most optimistic value quoted
by \citet{dessart_07_a} is only $1.5 \times 10^{-4}$ galactic events per
year.


\section{Results}
\label{sec:results}


\subsection{Dynamics of rotational stellar core collapse}
\label{sec:collapse_dynamics}

Previous simulations, considering a large variety of rotation rates
and profiles in the progenitor core but simplifying the complex
microphysics and/or the influence of GR, found qualitatively and
quantitatively different types of GW burst signals \citep[see e.g.\
the work by][]{moenchmeyer_91_a, zwerger_97_a, dimmelmeier_02_b}.
These can be classified depending on the collapse dynamics:
\emph{Type~I} signals are emitted when the collapse of the
homologously contracting inner core is not strongly influenced by
rotation, but stopped by a \emph{pressure-dominated bounce} due to the
stiffening of the EoS above nuclear matter density. This leads to the
formation of the PNS with a maximum core density
$ \rho_\mathrm{max} \ge \rho_\mathrm{nuc} $ after a few ring-down
oscillations. \emph{Type~II} signals occur when centrifugal forces,
which grow during contraction due to angular momentum conservation,
are sufficiently strong to halt the collapse, resulting in consecutive
(typically multiple) \emph{centrifugal bounces} with intermediate
coherent re-expansion of the inner core, seen as density drops by
often more than an order of magnitude; thus here
$ \rho_\mathrm{max} < \rho_\mathrm{nuc} $ after bounce.
\emph{Type~III} signals result from a pressure-dominated bounce when
the inner core has a very small mass at bounce due to a soft
subnuclear EoS or very efficient electron capture.

Accordingly, we discuss the quality of the new effective relativistic
potential by first applying it to rotating stellar cores which
collapse to form a PNS, focusing on three different cases: a regular
pressure-dominated bounce model with slow or moderate rotation, a
centrifugal single bounce model with rapid and strongly differential
rotation (characterized by a marked toroidal density
stratification\footnote{
In this paper, a configuration will be called ``toroidal'' if
the maximum density is reached off-center.
}
that develops in the core), and a multiple centrifugal bounce model.
As the Type~III bounce models have very similar dynamics to Type~I
models, we refrain from discussing them separately. The reader should
refer to Table \ref{tab:results_density} throughout the discussion,
as it presents key quantities (namely the maximum density at bounce
and the final maximum density) for our entire set of models, thus
providing further illustrative examples for moderately rotating
cores with weak (A1 models, s20A1B1) and strong (most A3 and A4 models)
differential rotation, as well as multiple (A2B4G1, A3B4G2) and single
(A4B5G5, A4B5G5 and s20A2B4) centrifugal bounce models.


\subsubsection{Slow and moderate rotation with pressure-dominated
  bounce}

\begin{figure}[t]
  \resizebox{\hsize}{!}
  {\includegraphics{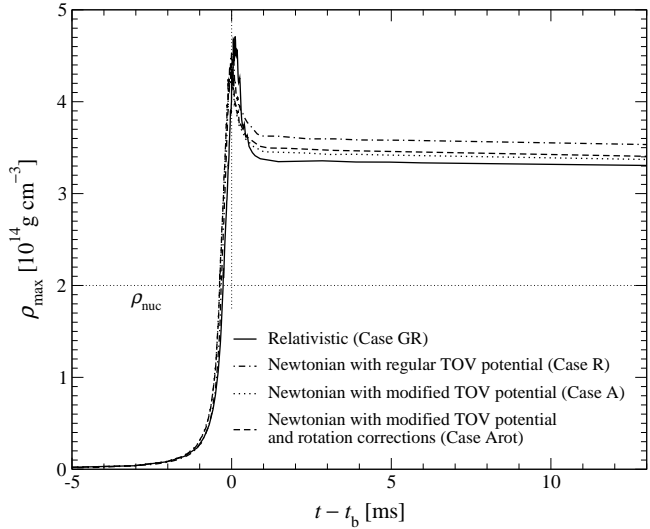}}
  \caption{Time evolution of the maximum density $ \rho_\mathrm{max} $
    for the slowly and almost uniformly rotating model A1B1G1 for
    Case~GR (solid line), Case~R (dashed-dotted line), Case~A (dotted
    line), and Case~Arot (dashed line), respectively. The results of
    all Newtonian simulations with effective relativistic potentials
    differ from the GR results by only a few percent. $ t_\mathrm{b} $
    is the time of core bounce.}
  \label{fig:slow_rotation}
\end{figure}

For the regular collapse models with slow or moderate rotation we
essentially confirm the findings of Paper~I, as the effective
relativistic potentials $ \Phi_\mathrm{Arot} $ and $ \Phi_\mathrm{A} $
are identical in the spherical limit. For the models A1B3G3 and A4B1G2
presented in Paper~I, as well as for all our other models with
the simple hybrid EoS which undergo pressure-dominated bounce,
$ \Phi_\mathrm{Arot} $ yields results that are in very good agreement
with those of GR simulations. In particular, after bounce and
ring-down $ \rho_\mathrm{max} $ is reproduced to an accuracy of
$ 7\% $ or better (see Table~\ref{tab:results_density}). Note that for
these nine models
${(T / |W|)}_\mathrm{b} < 0.1 $.
The maximum density
$ \rho_\mathrm{max,b} $ at bounce is usually a bit underestimated, but
correct to within $ 10\% $. Typically, Case~A matches Case~GR most
closely in the post-bounce phase, followed by Case~Arot (the new
potential including rotation corrections) and Case~R (the potential
with pure TOV corrections), while Case~Arot works best during
bounce. This behaviour is illustrated for model A1B1G1 in
Fig.~\ref{fig:slow_rotation}. The agreement is even better for the
microphysical model s20A1B1, whose initial rotation rate and profile
is identical to that of model A1B1G1. Here the error in
$ \rho_\mathrm{max} $ (compared to the GR case) is smaller than
$ 5\% $ at bounce (where now Case~R agrees best with Case~GR) and
$ 3\% $ after ring-down, as summarized in
Table~\ref{tab:results_density}.


\subsubsection{Multiple centrifugal bounce}
\label{subsubsec:multiple_centrifugal_bounce}

The limits in applying the old effective relativistic potentials
presented in Paper~I (Cases~A and~R) become apparent for models
which collapses slowly due to $ \gamma_1 $ being close to $ 4 / 3 $
and are strongly influenced by centrifugal forces due to a considerable
amount of initial rotation. Such models undergo multiple centrifugal
bounces where always $ \rho_\mathrm{max} < \rho_\mathrm{nuc} $,
and do not settle down to a quasi-equilibrium state even after
several bounces, and the resulting waveform of the
emitted gravitational radiation is of Type~II. It should be emphasized
that multiple bounces do no longer occur once deleptonization is taken
into account, essentially because a choice of $\gamma_1$ very close to
$ 4 / 3 $ does not constitute an adequate approximation for the dynamics
of the collapse. However, a rather high value of $\gamma_1$, such as to
produce multiple bounces, still serves as an interesting test case
which allows us to check the corrections due to rotation included for
Case~Arot in a regime where centrifugal forces dominate the evolution
of the core around bounce. For this reason, we consider one model
(A2B4G1) which provides a good example for the multiple bounce scenario.

\begin{figure}[t]
  \resizebox{\hsize}{!}
  {\includegraphics{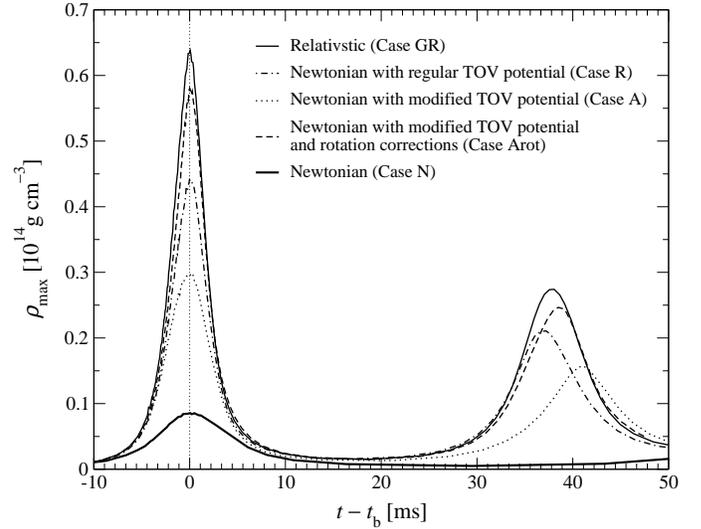}}
  \caption{Same as Fig.~\ref{fig:slow_rotation} but for the multiple
    centrifugal bounce model A2B4G1. The effective relativistic
    potential $ \Phi_\mathrm{Arot} $ reproduces the GR result to
    within $ 10\% $ even though this model is strongly influenced by
    centrifugal forces. The additionally shown purely Newtonian
    simulation grossly underestimates $ \rho_\mathrm{max} $.}
  \label{fig:multiple_centrifugal_bounce}
\end{figure}

In this model (and comparable models), $ \rho_\mathrm{max,b} $ at the
first bounce is underestimated by more than $ 50\% $ in Case~A and by
more than $ 30\% $ in Case~R (see
Fig.~\ref{fig:multiple_centrifugal_bounce}, and also Fig.~13 in
Paper~I), while the use of the new potential $ \Phi_\mathrm{Arot} $
reduces this error to less than $ 10\% $. This is an extraordinary
result, as in a collapse situation with multiple centrifugal bounces
the evolution of $ \rho_\mathrm{max} $ sensitively depends on the
balance between centrifugal forces and the pressure gradient for an
EoS with a $ \gamma_1 $ close to $ 4 / 3 $.

For rotational core collapse simulations with a simple hybrid EoS, the
appropriate modeling of GR effects (via a consistent relativistic
hydrodynamic formulation like in Case~GR or by the use of the new
effective relativistic potential $ \Phi_\mathrm{Arot} $ in an
otherwise Newtonian framework) strongly reduces the range in parameter
space where multiple centrifugal bounce models occur
\citep{dimmelmeier_02_b}. Moreover, independent of the inclusion of
relativistic gravity, only models with $ \gamma_1 \ge 1.31 $ will
exhibit such a bounce behavior, even in a purely Newtonian treatment.
However, as will be explained in detail in the next
Sect.~\ref{subsubsec:single_centrifugal_bounce}, the occurence of
models showing multiple centrifugal bounces results from the use of a
simplified EoS, and vanishes if an appropriate microphysical matter
model is applied.

\subsubsection{Rapid and strongly differential rotation with single
  centrifugal bounce}
\label{subsubsec:single_centrifugal_bounce}

If the microphysical EoS and the approximate deleptonization scheme
described in Sect.~\ref{subsec:models} are employed, the collapse
dynamics and the GW burst signature are exclusively of Type~I. This
was first observed by \citet{ott_07_a}, and then confirmed by
\citet{dimmelmeier_07_a} who considerably extended the number of
investigated models and comprehensively explored a wide parameter
space of initial rotation states. They found that the combination of
deleptonization during core collapse and (to a lesser degree) the
influence of GR lead to this emergence of a generic GW signal type. In
most microphysical simulations, the core undergoes a regular
pressure-dominated bounce, which results in a dynamical behavior that
is well represented by the evolution of $ \rho_\mathrm{max} $ shown in
Fig.~\ref{fig:slow_rotation}. Accordingly, the new effective
relativistic potential $ \Phi_\mathrm{Arot} $ as well as potentials
$ \Phi_\mathrm{A} $ and $ \Phi_\mathrm{R} $ again yield excellent
results, as they also do for models based on a simple hybrid EoS (see
model s20A1B1 in Table~\ref{tab:results_density}).

\begin{figure}
  \resizebox{\hsize}{!}
  {\includegraphics{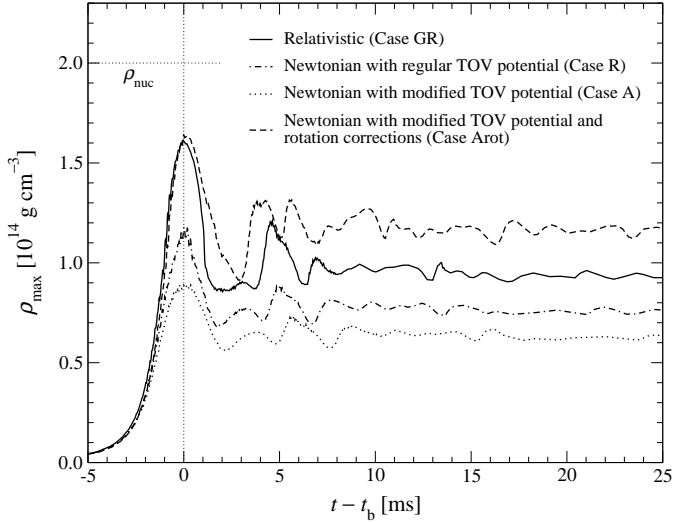}}
  \caption{Same as Fig.~\ref{fig:slow_rotation} but for the
    centrifugally bouncing microphysical model s20A2B4. The effective
    relativistic potential $ \Phi_\mathrm{Arot} $ yields very good
    results around core bounce, but later causes an overestimation of
    $ \rho_\mathrm{max} $. In a purely Newtonian simulation this model
    does not even start to contract.}
  \label{fig:microphysical_model}
\end{figure}

Only for models with very fast initial rotation (with an
initial value\footnote{
Of course, the precise value for the initial rotation parameter
beyond which a centrifugal bounce at subnuclear densities occurs also
depends on the rotation profile and the choice of $\gamma_1$.
\citet{dimmelmeier_07_a} should be referred to for a detailed analysis.
}
of $T/|W| \gtrsim 0.01\ldots 0.02$), the core bounce is
not caused by the stiffening of the EoS but rather by centrifugal
forces.
However, in contrast to the multiple centrifugal bounce
behavior of the (simple hybrid EoS) model A2B4G1 discussed in
Sect.~\ref{subsubsec:multiple_centrifugal_bounce}, with microphysics
only a single bounce with subsequent formation of a quasi-stationary
PNS is observed, reflected by a Type~I waveform. 
For such cases, where rotation plays a similarly crucial role for
the dynamics as e.g.\ during the disk formation phase in the collapsar
model, the potential $ \Phi_\mathrm{Arot} $ now clearly outperforms
potentials $ \Phi_\mathrm{A} $ and $ \Phi_\mathrm{R} $ (see
Fig.~\ref{fig:microphysical_model}). 
At core bounce we find
$ \rho_\mathrm{max,b} = 1.64 \times 10^{14} \mathrm{\ g\ cm}^{-3} $
for Case~Arot, which is only $ 2\% $ larger than the value
$ 1.61 \times 10^{14} \mathrm{\ g\ cm}^{-3} $ obtained in the GR
simulation, while in Cases~A and~R the value of
$ \rho_\mathrm{max,b} $ is underestimated by $ 27\% $, and $ 45\% $,
respectively. In the post-bounce phase, however,
$ \Phi_\mathrm{Arot} $ leads to an overestimatate of
$ \rho_\mathrm{max,b} $ by roughly the same amount as
$ \Phi_\mathrm{R} $ underestimates the correct value. In summary, for
this rapidly rotating configuration of a core that is stabilized
against gravity by centrifugal forces rather than pressure gradients,
the new effective relativistic potential with rotational corrections
performs better than the old potentials at bounce and at least as good
after bounce.

We also emphasize that the microphysical model s20A2B4 does not even
start to contract in Newtonian gravity due to rotational
stabilization. This case where a GR instability leads to the
contraction of a core that is still well in the Newtonian limit,
clearly illustrates the usefulness of the effective relativistic
potential approach. Such a GR instability is also responsible for the
collapse of supermassive stars with masses between $ 10^4 $ and
$ 10^8 \, M_\odot $, which are candidates for being progenitors of
supermassive black holes. Our results suggest that in this scenario
the use of an effective relativistic potential should be able to
capture the qualitatively correct collapse dynamics, too, while a
purely Newtonian treatment fails.

\begin{figure}[t]
  \resizebox{\hsize}{!}
  {\includegraphics{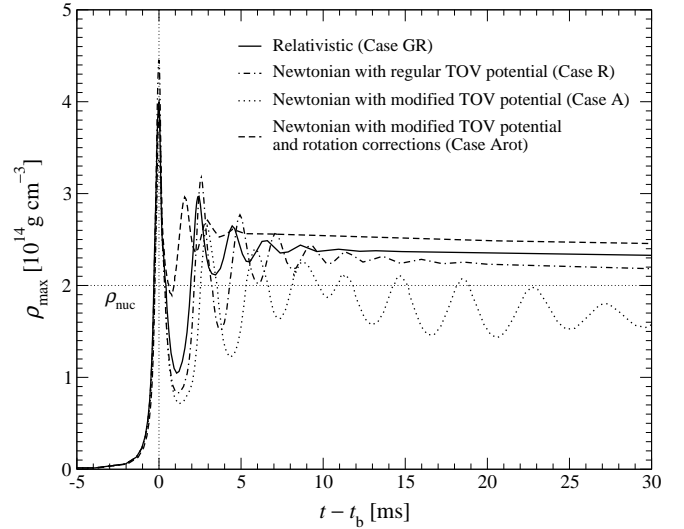}}
  \caption{Same as Fig.~\ref{fig:slow_rotation} but for the rapidly
    and strongly differentially rotating model A4B2G3. Unlike the
    effective relativistic potentials $ \Phi_\mathrm{TOV} $ and $
    \Phi_\mathrm{Arot} $, potential $ \Phi_\mathrm{A} $ cannot
    reproduce the correct (supranuclear) maximum density of the core
    after bounce.}
  \label{fig:strong_rotation}
\end{figure}

\begin{figure}[t]
  \resizebox{\hsize}{!}
  {\includegraphics{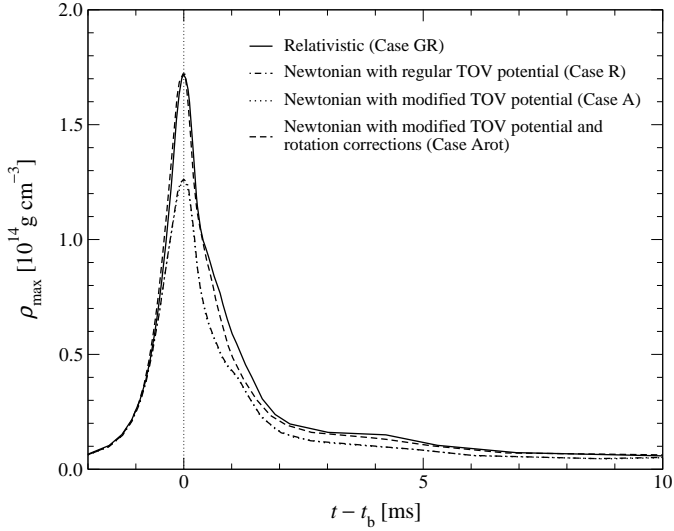}}
  \caption{Same as Fig.~\ref{fig:slow_rotation} but for the single
    centrifugal bounce model A4B5G5. Again the results obtained with
    the effective relativistic potential $ \Phi_\mathrm{Arot} $ show
    excellent agreement with those found in GR. Note that in this
    model the density maximum is located off-center, and that the
    differences between the Case~A and Case~R are minimal.}
  \label{fig:single_centrifugal_bounce}
\end{figure}

If a simple hybrid EoS is used, the situation which comes closest to
that of the single centrifugal bounce of the microphysical models is
when fast initial rotation is combined with a subnuclear adiabatic
index $ \gamma_1 $ that is significantly lower than $ 4 / 3 $. Then
the collapse of the core is also stopped by only a single bounce
(albeit with possibly very strong ring-down oscillations as seen for
model A4B2G3 in Fig.~\ref{fig:strong_rotation}, or a considerable
re-expansion of the core by sometimes more than a factor of 10 in
$ \rho_\mathrm{max} $ as shown for model A4B5G5 in
Fig.~\ref{fig:single_centrifugal_bounce}). The post-bounce PNS is
either exclusively or predominantly stabilized by centrifugal
forces. Many such models have a toroidal density stratification where
the off-center maximum density $ \rho_\mathrm{max} $ exceeds the
central density $ \rho_\mathrm{c} $ by more than one order of
magnitude.

In such models the failure of an effective relativistic potential
without corrections due to rotation becomes particularly apparent when
only a small (possibly off-center) region of the core reaches
supranuclear densities and the bounce is caused by a combination of
the stiffening of the EoS and by centrifugal forces, as in model
A4B2G3 (see Fig.~\ref{fig:strong_rotation}). Here, the potential
$ \Phi_\mathrm{A} $ is unable to produce a PNS that is sufficiently
compact for the maximum density to remain above $ \rho_\mathrm{nuc} $,
resulting in a post-bounce configuration where $ \rho_\mathrm{max} $
is $ 33\% $ lower than in the GR case. The rotational corrections
incorporated into the potential $ \Phi_\mathrm{Arot} $ on the other
hand eliminate this weakness and lead to a final $ \rho_\mathrm{max} $
which is correct to within $ 6\% $.

The results presented in Figs.~\ref{fig:strong_rotation}
and~\ref{fig:single_centrifugal_bounce} demonstrate that for very
rapidly rotating models with a simple matter prescription, the new
potential $ \Phi_\mathrm{Arot} $ is superior to $ \Phi_\mathrm{A} $
both at bounce (showing perfect agreement with the GR results for
model A4B5G5) and in the post-bounce phase, while it performs at least
equally well as $ \Phi_\mathrm{R} $. We again emphasize that the TOV
potential $ \Phi_\mathrm{R} $ does not give the correct results for
slow or vanishing rotation, and is thus not as versatile in handling a
wide variety of rotation states as $ \Phi_\mathrm{Arot} $.

In summary, we find that unlike the other effective relativistic
potentials, $ \Phi_\mathrm{Arot} $ reproduces the GR results to within
an error of at most $ 15\% $ in $ \rho_\mathrm{max} $ in the
\emph{entire} investigated parameter space of rotation for both simple
and microphysical models. The new effective relativistic potential
also neither produces an overly compact PNS in spherical symmetry or
for slow rotation -- like the TOV potential $ \Phi_\mathrm{R} $ -- nor
does it fail for centrifugal bounces as does $ \Phi_\mathrm{A} $, and
to a smaller extent also $ \Phi_\mathrm{R} $.


\subsection{Structure and deformation of proto-neutron stars}
\label{sec:pns_structure}

Following Paper~I, we do not confine our investigation to local
quantities like the maximum density $ \rho_\mathrm{max} $, but also
study the spatial structure of each model by comparing radial profiles
obtained in GR with simulations using the new effective relativistic
potential with corrections due to rotation.

All effective relativistic potentials discussed in this work as well
as in Paper~I are constructed from a solution of the TOV equation,
which is formulated in Schwarzschild radial coordinates. Therefore,
our (otherwise) Newtonian simulations presuppose a specific gauge
choice which is different from that used in the GR simulations, where
the metric equations in the CFC approximation are based on isotropic
radial coordinates. As a consequence, it is necessary to apply a
coordinate transformation in order to be able to perform a resonable
comparison between both classes of models.

As in Paper~I we take the standard Schwarzschild line element in
spherical symmetry,
\begin{equation}
  \mathrm{d}s^2 = - \alpha^2 \, \mathrm{d}t^2 +
  \frac{1}{1 - 2 m / r} \, \mathrm{d}r^2 + r^2 \, \mathrm{d}\Omega^2,
\end{equation}
and replace it by the one in isotropic radial coordinates:
\begin{equation}
  \mathrm{d}s_\mathrm{iso}^2 = - \alpha^2 \, \mathrm{d}t^2 +
  \phi^4 \left( \mathrm{d}r_\mathrm{iso}^2 +
  r_\mathrm{iso}^2 \, \mathrm{d}\Omega^2 \right),
\end{equation}
For details about this transformation we refer to Appendix~B of
Paper~I, in particular Eq.~(B.8).

\begin{figure}
  \resizebox{\hsize}{!}
  {\includegraphics{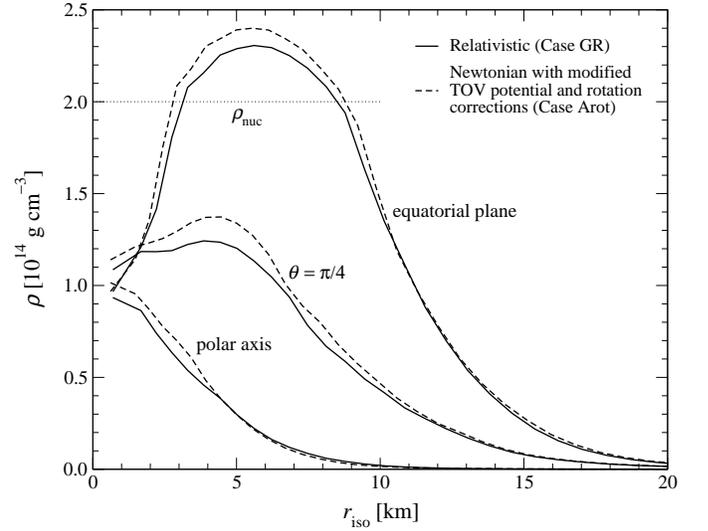}}
  \\ [1.0 em]
  \resizebox{\hsize}{!}
  {\includegraphics{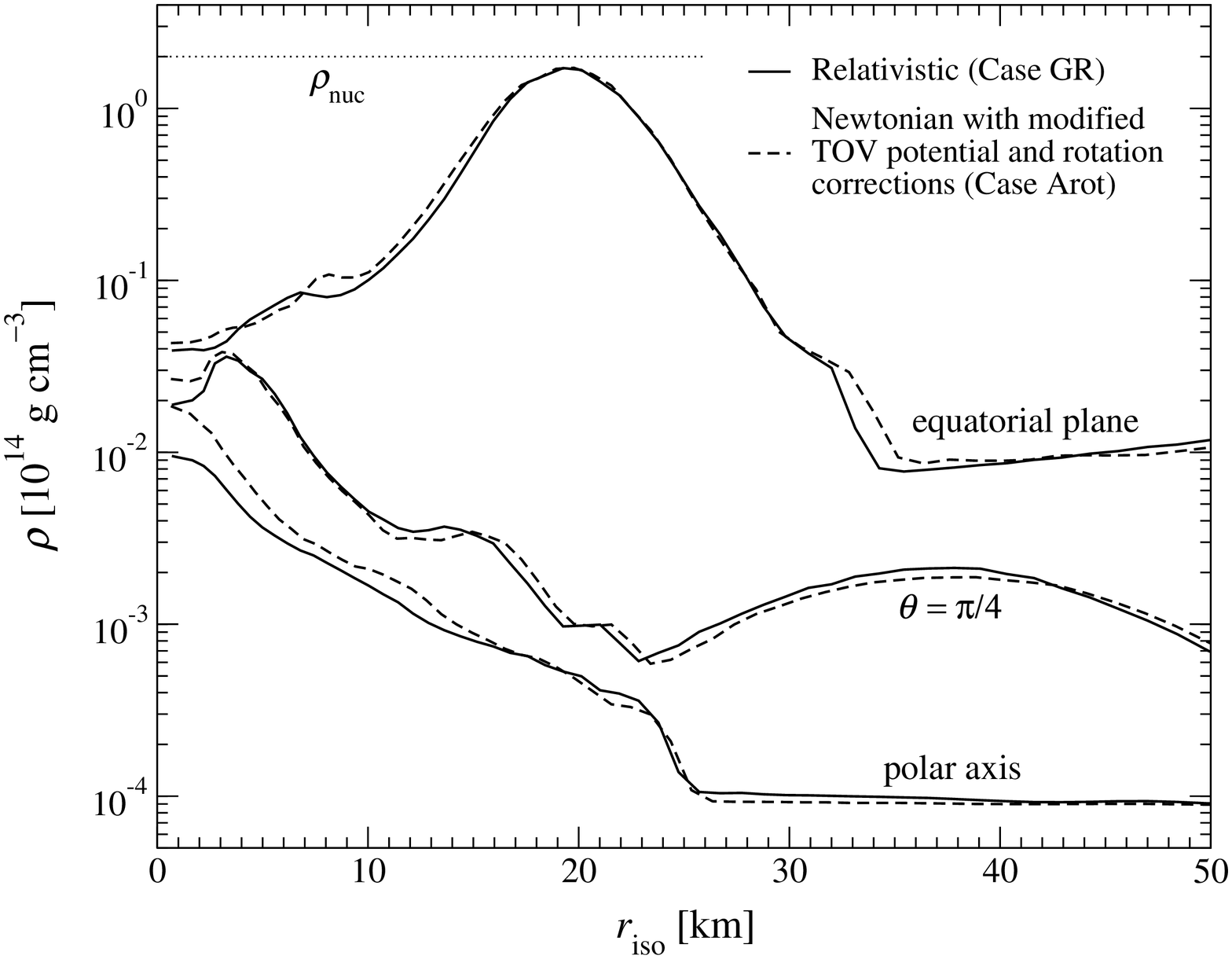}}
  \caption{Radial profiles of the density $ \rho $ for model A4B2G3
    (top panel; $ 40 \mathrm{\ ms} $ after bounce) and model A4B5G5
    (bottom panel, at the time of bounce) at various polar angles for
    Case~GR (solid lines) and Case~Arot (dashed lines), respectively.
    For the potential $ \Phi_\mathrm{Arot} $ we apply a transformation
    to the isotropic radial coordinate (see text for details). Both
    models rotate rapidly and strongly differentially, and model
    A4B5B5 bounces at subnuclear densities.}
  \label{fig:radial_profiles_density}
\end{figure}

\begin{figure}
  \resizebox{\hsize}{!}
  {\includegraphics{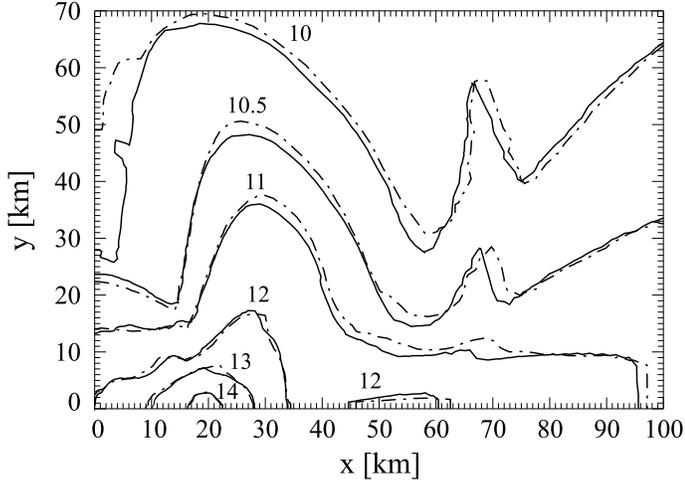}}
  \caption{
    Density contours for model A4B5G5 at bounce in full GR
    (solid lines), and for Case~Arot (dashed-dotted lines).  $\log_{10} \rho$
    is indicated by the contour labels. As in
    Fig.~\ref{fig:radial_profiles_density} we apply a transformation to
    isotropic radial coordinates.}
  \label{fig:density_contours}
\end{figure}

\begin{figure}[t]
  \resizebox{\hsize}{!}
  {\includegraphics{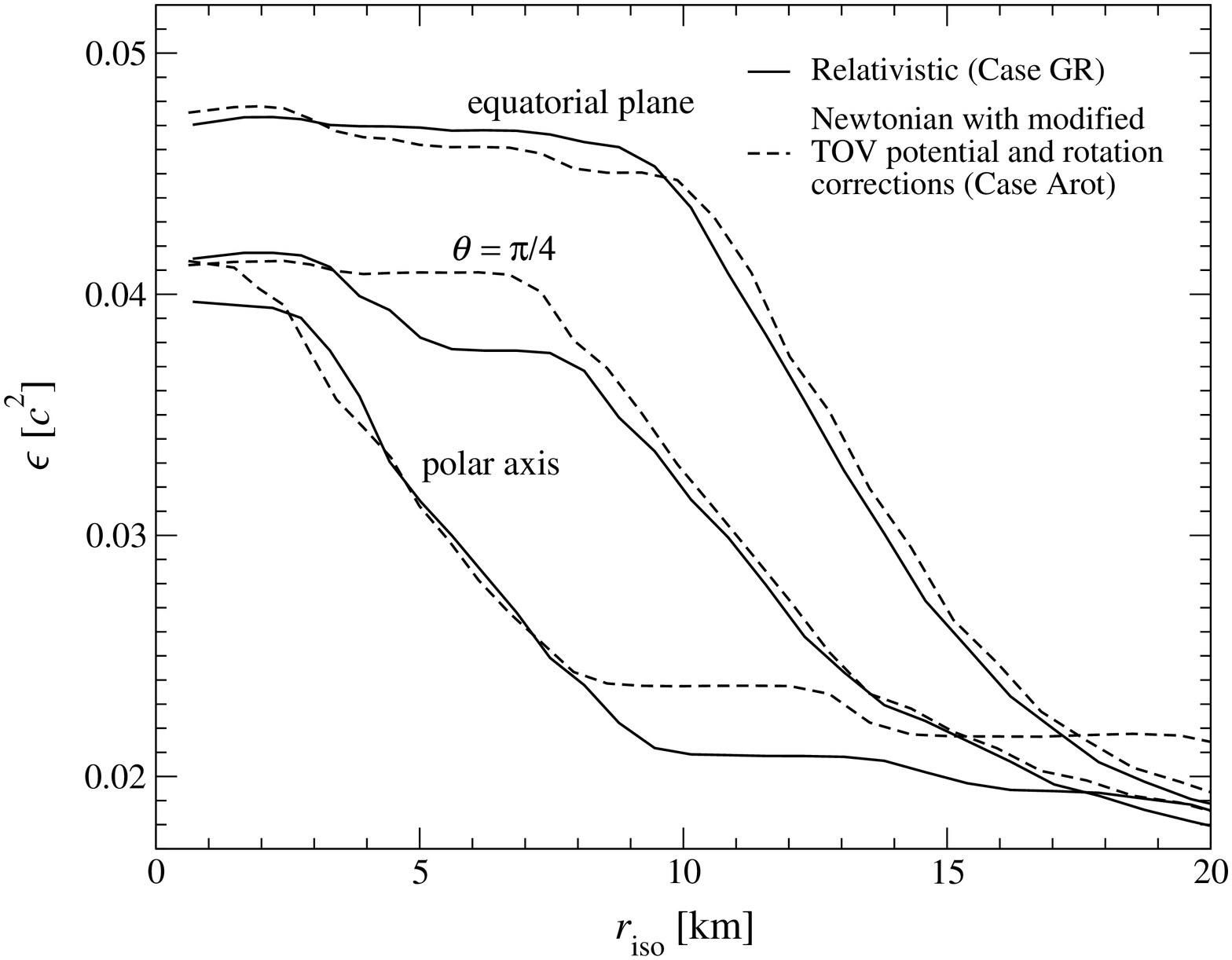}}
  \\ [1.0 em]
  \resizebox{\hsize}{!}
  {\includegraphics{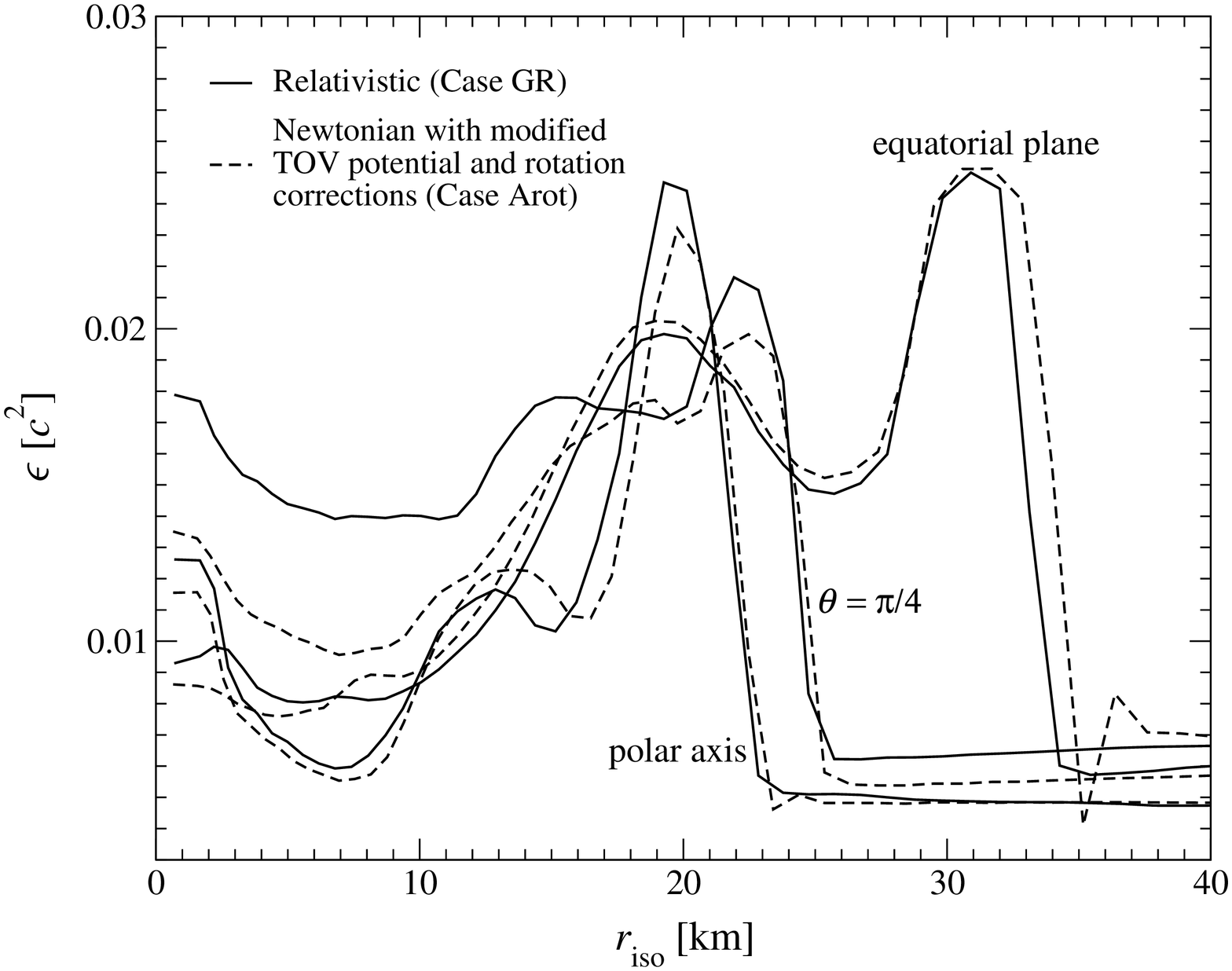}}
  \caption{Same as Fig.~\ref{fig:radial_profiles_density}, but for the
    specific internal energy $ \epsilon $. Note that potential
    $ \Phi_\mathrm{Arot} $ (dashed lines) closely reproduces the
    step-like features in $ \epsilon $ of the model in GR (solid
    lines).}
  \label{fig:radial_profiles_internal_energy}
\end{figure}

As long as the shift vector $ \beta^i $ is close to zero and the
deviation of the metric functions $ \alpha $ and $ \phi $ from
spherical symmetry are not too large, $ r_\mathrm{iso} $ then yields a
good estimate for the (in general not spherically symmetric) GR
spacetime metric. These conditions are very well satisfied even during
the most dynamic phases of the collapse: The radial shift vector
component typically reaches a maximum value of the order
$ |\beta_r| \approx 0.01 $ (in units of the speed of light $ c $) near
bounce and then decreases as the PNS settles down to a
quasi-stationary state, while the meridional component
$ \beta_\theta $ is even smaller. The rotational component
$ \beta_\varphi $ is about one order of magnitude smaller than the
maximum rotational velocity $ v_\varphi $, i.e.\ its absolute value
rarely exceeds $ 0.01 $ even for the most rapidly rotating models.
Moreover, the anisotropy of the metric functions $ \alpha $ and
$ \phi $ is significantly smaller than the anisotropy of the matter
fields.

The potentials $ \Phi_\mathrm{A} $ and $ \Phi_\mathrm{Arot} $ are
identical in the non-rotating limit and differ only slightly for slow
rotation, as the corrections due to rotation are negligible. Thus, we
concentrate our discussion on the two rapidly rotating models A4B2G3
and A4B5G5. Furthermore, we investigate only the innermost regions of
the collapsed stellar core constituting the nascent PNS, which is in
approximate equilibrium after the core bounce and the ring-down phase.

Figs.~\ref{fig:radial_profiles_density}
and~\ref{fig:radial_profiles_internal_energy} illustrate that the
radial profiles of the density $ \rho $ and the specific internal
energy $ \epsilon $ in Case~Arot are in very good agreement with those
obtained in GR for both models. To demonstrate that this excellent
matching is independent of the angular location, we show profiles of
$ \rho $ at three different latitudes -- along the polar axis, at
$ \theta = \pi / 4 $, and in the equatorial plane, respectively. For
model A4B2G3, in both cases (Arot and GR) we observe after the
ring-down phase a similar toroidal structure with a maximum density of
$ \rho_\mathrm{max} \approx (2.3\mbox{\,--\,}2.4) \times 10^{14}
\mathrm{\ g\ cm}^{-3} $ at a radius of
$ 5\mbox{\,--\,}6 \mathrm{\ km} $ in the equatorial plane, and a
central density $ \rho_\mathrm{c} $ close to
$ 1.0 \times 10^{14} \mathrm{\ g\ cm}^{-3} $. 
Even for model A4B5G5,
which develops a very extreme toroidal structure and is depicted in
the highly dynamic phase during core bounce
in the lower panel of Fig.~\ref{fig:radial_profiles_density}, the new
potential $ \Phi_\mathrm{Arot} $ yields results which closely match those in
GR. The maximum density $ \rho_\mathrm{max} $ is reached at
$ r \approx 20 \mathrm{\ km} $ in the equatorial plane. 
Fig.~\ref{fig:density_contours} shows density contours for
this model down to densities of $10^{10} \mathrm{\ g\ cm}^{-3}$,
and demonstrates how closely Case~Arot reproduces the spatial
structure of that model in full GR. The step-like features in
$ \epsilon $ in the low-density regions, which
are caused by shock heating, are also reproduced quite accurately,
as can be seen in Fig. \ref{fig:radial_profiles_internal_energy}.

Due to the very good agreement of the density and internal energy
profiles between Cases~GR and~Arot already during the dynamical phase
of the core bounce we also infer that the use of the transformation
from Schwarzschild to isotropic radial coordinates needs not be
limited to the case discussed in Paper~I, i.e.\ a spherical PNS in the
late post-bounce phase, but works even for strongly rotating and
significantly aspherical configurations.


\subsection{Structure of rotating neutron stars}

\begin{figure}
  \resizebox{\hsize}{!}
  {\includegraphics{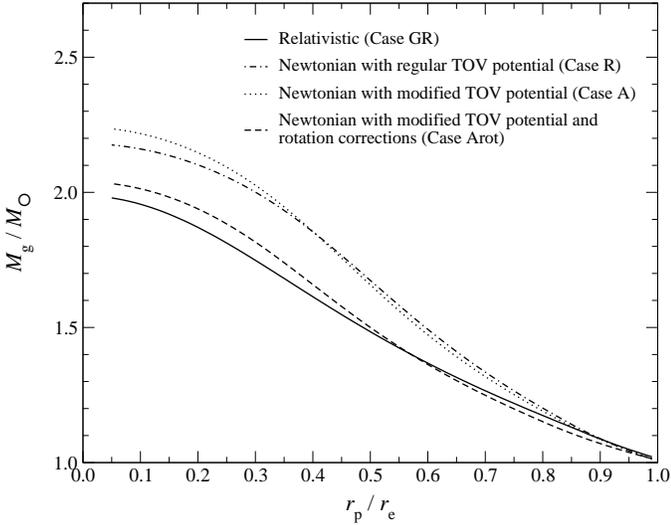}}
  \caption{Dependence of the gravitational mass $ M_\mathrm{g} $ on
    the axis ratio $ r_\mathrm{p} / r_\mathrm{e} $ for rotating NS
    models with a maximum density $ \rho_\mathrm{max} = 3.95 \times
    10^{14} \mathrm{\ g\ cm}^{-3} $, and a differential rotation
    parameter $ \hat {A} = 0.3 $. We show the results for Case~GR
    (solid line), Case~R (dashed-dotted line), Case~A (dotted line),
    and Case~Arot (dashed line), respectively. The results obtained
    with the effective relativistic potential $ \Phi_\mathrm{Arot} $
    including corrections due to rotation agree best with the ones of
    GR.}
  \label{fig:mass_axis_ratio_relation}
\end{figure}

The core collapse simulations presented here are limited to a maximum
compactness of the PNS of
$ 2 M_\mathrm{PNS} / R_\mathrm{PNS} \sim 0.2 $ in the early
post-bounce phase, where $ M_\mathrm{PNS} $ and $ R_\mathrm{PNS} $ are
the mass and radius of the PNS, respectively. In order to extend the
assessment of the quality of a Newtonian simulation with an effective
relativistic potential to a more strongly relativistic regime, we also
investigate equilibrium models of a rotating NS using the polytropic
EoS~(\ref{eq:polytrope}). These can serve as an approximation to
either a cooling PNS which has already evolved over several
$ 100 \mathrm{\ ms} $ (and cannot be directly simulated with our code
due to computational constraints and the missing necessary
microphysics) or a cold NS. Such models also enable us to check the
reliability of our approach for scenarios such as accretion-induced
collapse. Similarly, the ability to correctly handle equilibrium
models of compact objects with the effective relativistic potential
approach is a prerequisite for investigating instabilities or
oscillations in a NS.

Therefore, we also perform an analysis of various rotating NS models
in GR and for the different effective relativistic potentials. We find
that the new potential $ \Phi_\mathrm{Arot} $ reproduces quite closely
the correct GR dependence of most global quantities (like
gravitational mass, binding energy, etc.) on e.g.\ the axis ratio of
the NS, even for strong and differential rotation, and significant
toroidal deformation.

As long as the NS is not too compact (i.e.\
$ \rho_\mathrm{c} < 4 \times 10^{14} \mathrm{\ g \ cm}^{-3} $, which
are typical values for a PNS during the first few
$ 100 \mathrm{\ ms} $ after core bounce), the results for Case~R and~A
do not differ strongly. This is evident from
Fig.~\ref{fig:mass_axis_ratio_relation}, where we plot the
gravitational mass $ M_\mathrm{g} $ for a sequence of NS models with
constant central density
$ \rho_\mathrm{c} = 3.95 \times 10^{14} \mathrm{\ g\ cm}^{-3} $ and
differential rotation parameter $ \hat{A} = 0.3 $ (where
$ \hat{A} = A / r_\mathrm{e} $), and vary the axis ratio
$ r_\mathrm{p} / r_\mathrm{e} $. Here $ r_\mathrm{p}$ and
$ r_\mathrm{e} $ are the polar and equatorial radius of the NS,
respectively. Fig.~\ref{fig:mass_axis_ratio_relation} shows that even
for an extreme axis ratio $ r_\mathrm{p} / r_\mathrm{e} = 0.05 $, the
deviation of $ M_\mathrm{g} $ from its GR value is only about $ 15\% $
for the worst potential (Case~A), and much better for potential
$ \Phi_\mathrm{Arot} $. 
The gravitational binding energy $ W $ is somewhat more sensitive to
the inclusion of relativistic effects; we obtain a considerably larger
value of $ |W| = 6.01 \times 10^{53} \mathrm{\ erg} $ for Case~A
compared to $ |W| = 4.60 \times 10^{53} \mathrm{\ erg} $ in full GR
($ +31\% $). However, using the potentials $ \Phi_\mathrm{TOV} $ or
$ \Phi_\mathrm{Arot} $ we obtain
$ |W| = 5.31 \times 10^{53} \mathrm{\ erg} $ ($ +15\% $)
and $ |W| = 5.04 \times 10^{53} \mathrm{\ erg} $ ($ +10\% $), which is
reasonably close to the GR value.
We also find good agreement for quantities
like the baryonic mass, and the central angular
velocity, as well as for the spatial structure of the NS models (e.g.\
polar and equatorial radii, density profiles).

\begin{figure}
  \resizebox{\hsize}{!}
  {\includegraphics{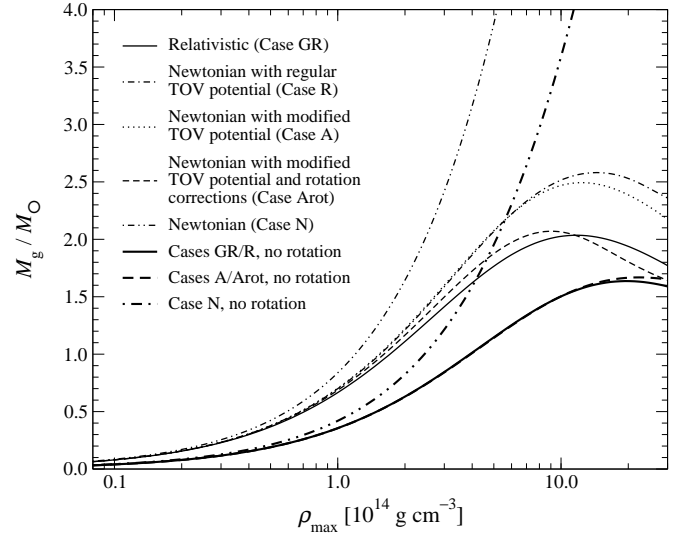}}
  \caption{Same as Fig.~\ref{fig:mass_axis_ratio_relation}, but
    showing the dependence of the gravitational mass $ M_\mathrm{g} $
    on the maximum mass $ \rho_\mathrm{max} $ for rotating NS models
    with an axis ratio $ r_\mathrm{p} / r_\mathrm{e} = 0.4 $ and a
    differential rotation parameter $ \hat {A} = 0.3 $. Again the
    results obtained with the effective relativistic potential
    $ \Phi_\mathrm{Arot} $ agree best with those in GR, while in the
    purely Newtonian case the values for $ M_\mathrm{g} $ are
    consistently too high. Without rotation (thick lines) the Cases GR
    and R, as well as A and Arot are identical, respectively.}
  \label{fig:mass_density_relation}
\end{figure}

For more compact configurations, we find the deviations from the GR
models to be more pronounced, as illustrated by
Fig.~\ref{fig:mass_density_relation}, where we show $ M_\mathrm{g} $
as a function of the maximum density $ \rho_\mathrm{max} $ for
$ \hat{A} = 0.3 $ and an axis ratio
$ r_\mathrm{p} / r_\mathrm{e} = 0.4 $. Apparently the error in
$ M_\mathrm{g}$ reaches $ 0.6 \, M_\odot $ for potential
$ \Phi_\mathrm{A} $, and $ 0.47 \, M_\odot $ for potential
$ \Phi_\mathrm{TOV} $, but does not exceed $ 0.15 \, M_\odot $ for the
new potential $ \Phi_\mathrm{Arot} $ with rotational corrections.

It is also noteworthy that for fast rotation the deviations from GR in
Cases~A and~R already manifest themselves well below the density
$ \rho_\mathrm{max} \approx 5 \times 10^{14} \mathrm{\ g\ cm}^{-3} $
of maximum gravitational mass, which is the transition point to the
unstable branch. For non-rotating models the deviations remain small
even beyond that point, which is then located at
$ \rho_\mathrm{max} \approx 20 \times 10^{14} \mathrm{\ g\ cm}^{-3} $
(see Fig.~\ref{fig:mass_density_relation}, and cp.\ Paper~I). On the
other hand, the purely Newtonian results overestimate $ M_\mathrm{g} $
by a factor of $ 2 $ already for
$ \rho_\mathrm{max} = 8 \times 10^{14} \mathrm{\ g\ cm}^{-3} $,
regardless of the strength of rotation. Thus, the effective
relativistic potentials, and particularly the new one
$ \Phi_\mathrm{Arot} $ give a clear improvement over a purely
Newtonian treatment. As for the core collapse simulations, Case~Arot
is very robust and accurate over the entire range of rotation states
with varying $ \hat{A} $ and $ r_\mathrm{p} / r_\mathrm{e} $. It
exhibits a maximum error in $ M_\mathrm{g} $ of $ \approx 10\% $
compared to more than $ 20\% $ and $ 30\% $ for Case~R and Case~A,
respectively.

\subsection{Oscillations of neutron stars}

The effective relativistic potential approach obviously approximates
very well the collapse dynamics of rapidly rotating stellar cores as
compared to a simulation in GR. Furthermore, it also gives accurate
results for the structure of a nascent PNS after core collapse as well
as of a rotating cold NS. We now subject this approach to an even more
stringent test for its applicability to the post-bounce evolution of a
PNS or to instability problems in a rotating NS. To this end, we
consider the dynamics of a NS near equilibrium and oscillations in
the linear regime. For convenience, we restrict the study to
non-rotating models (where Case~A and~Arot are identical), as the
results discussed here are qualitatively independent of rotation.

\begin{figure}
  \resizebox{\hsize}{!}
  {\includegraphics{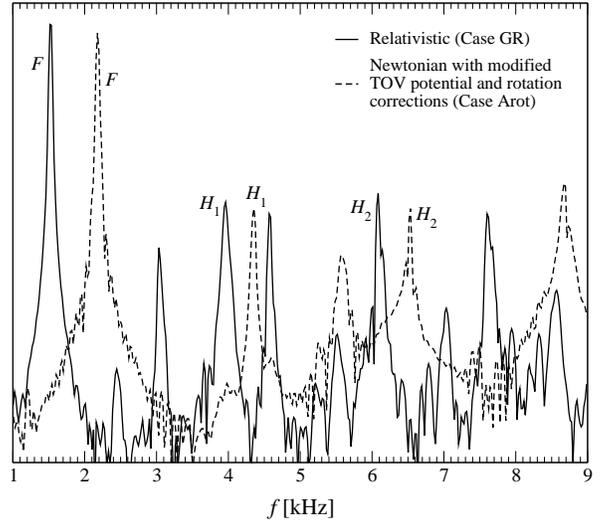}}
  \caption{Power spectrum of the time evolution of the maximum density
    $ \rho_\mathrm{max} $ for a very compact non-rotating NS model for
    Case~GR (solid line) and Case~Arot (dashed line), respectively.
    The $ F $-mode and its harmonics $ H_1 $ and $ H_2 $ are clearly
    visible. For all investigated modes the frequencies obtained with
    the potential $ \Phi_\mathrm{Arot} $ are significantly higher than
    those in GR.}
  \label{fig:power_spectrum}
\end{figure}

We first evolve a very compact polytropic $ \gamma = 2 $ NS model with
$ \rho_\mathrm{c} = 7.9 \times 10^{14} \mathrm{\ g\ cm}^{-3} $, which
is a density value typically reached around $ 1\mathrm{\ s} $ after
core bounce. When applying a small radial velocity perturbation to
excite radial normal modes, we find significantly different
frequencies of the eigenmodes as compared to GR if we use any of the
effective relativistic potentials discussed here. For instance, all
these potentials overestimate the frequency of the fundamental radial
$ F $-mode by about $ 45\% $. This can be clearly seen in
Fig.~\ref{fig:power_spectrum}, where we present the power spectrum of
the time evolution of $ \rho_\mathrm{max} $, in which the eigenmodes
can be identified as distinct peaks.

For a less compact configuration with
$ \rho_\mathrm{c} = 3.95 \times 10^{14} \mathrm{\ g\ cm}^{-3} $, which is obtained a
few $ 100 \mathrm{\ ms} $ after bounce, the error is somewhat smaller
($ \sim 22\% $) but still much larger than what could be expected
given the excellent agreement in the structure of the NS model between
the GR case and the use of an effective relativistic potential. In
addition, the higher harmonics of the $ F $-mode like the $ H_1 $-mode
or the $ H_2 $-mode also exhibit this apparent discrepancy (see
Fig.~\ref{fig:power_spectrum}).

The failure of the effective relativistic potential method in the
context of NS oscillations can be understood by comparing the
equations of hydrodynamics in GR and the Newtonian
formulation\footnote{We remind the reader here 
that except for
  changes in the gravitational potential source terms, in the
  effective relativistic potential approach the equations of Newtonian
  hydrodynamics remain unaltered.}. Limiting ourselves to the most
straightforward case of the TOV potential in Case~R and to spherical
symmetry, we find the following evolution equation for the radial
velocity:
\begin{equation}
  \frac{\pd v_r}{\pd t} = - \frac{1}{\rho} \frac{\pd P}{\pd r} -
  \frac{\pd \: \phicapbar_\mathrm{TOV}}{\pd r}.
  \label{eq:ns_oscillation_newtonian}
\end{equation}
An analogous equation can be obtained in GR, starting from the
formulation of \citet{vanriper_79_a}, which uses Schwarzschild
coordinates like in the case of the TOV potential. After rearranging
some terms to allow direct identifications with Case~R, one arrives at
\begin{equation}
  \frac{\pd \left( \alpha^{-1} v_r \right)}{\pd t} =
  \frac{\alpha}{h} \left( - \frac{1}{\rho} \frac{\pd P}{\pd r} -
  \frac{\pd \: \phicapbar_\mathrm{TOV}}{\pd r} \right).
  \label{eq:ns_oscillation_gr}
\end{equation}
Unlike the Newtonian equation~(\ref{eq:ns_oscillation_newtonian}), the
GR equation~(\ref{eq:ns_oscillation_gr}) contains an additional factor
given by the ratio of the lapse function $ \alpha $ and the specific
relativistic enthalpy $ h $. For a stationary state, the left hand
sides vanish and both equations become equivalent. This explains why
potential $ \Phi_\mathrm{TOV} $ (or other modified potentials like
$ \Phi_\mathrm{A} $ and $ \Phi_\mathrm{Arot} $) can be successfully
applied to model the structure of PNS or NS models.

On the other hand, in the general dynamic case, the additional factor
cannot be discarded, and will, for example, be present in the
linearized perturbation equations. Since the lapse function $ \alpha $
is significantly below $ 1 $ in compact NS (with a central value of
$ \alpha_\mathrm{c} \approx 0.7 $), this factor has a considerable
impact on the linearized perturbation equations and consequently on
the eigenfrequency spectrum. For special cases, such as spherical
pulsations, it may still be possible to obtain simple scaling
relations for some of the effective relativistic potentials, but this
is beyond the scope of this work. At present, however, we conclude
that the effective relativistic potential approach is much less suited
for studying the frequencies of NS oscillations than for approximating
e.g.\ the structure or global quantities like mass and radius. 
It is
conceivable that a pseudo-Newtonian treatment should be used with caution
whenever NS or PNS oscillations play a crucial dynamical role, e.g.\
in the acoustic explosion mechanism suggested by \citet{burrows_06_a},
whose potential is still disputed \citep{weinberg_08_a}. \emph{A fortiori},
this also applies for a purely Newtonian approach. A fully relativistic
treatment of the hydrodynamics would probably be desirable
in such a context to obtain firm quantitative predictions, even though
it may not be a necessary ingredient for a qualitatively correct model.

\subsection{Improvement of the gravitational wave extraction}
\label{sec:wave_signals}

After having investigated and compared the quality of the new
potential $ \Phi_\mathrm{Arot} $, we now focus on the issue of GW
emission from simulations of rotational stellar core collapse using
the effective relativistic potential approach. In our simulations, we
compute the radiation quadrupole moment $ A_{20}^\mathrm{E2} $
according to the first time-integrated Newtonian quadrupole formula
\citep{finn_89_a, blanchet_90_a},
\begin{eqnarray}
  A_{20}^\mathrm{E2} & = & \frac{32 \pi^{3/2}}{\sqrt{15}}
  \frac{\mathrm{d}}{\mathrm{d} t} \int r^3 \sin \theta \,
  \mathrm{d} \theta \, \mathrm{d} r \cdot
  \label{eq:quadrupole_formula}
  \\ [-0.7 em]
  & & \qquad \qquad \quad ~~
  \rho \,
  [v_r (3 \cos^2 \theta - 1) - 3 v_\theta \, \sin \theta \, \cos
  \theta].
  \nonumber
\end{eqnarray}%
We restrict the integration to those grid cells separated from the
initial low-density atmosphere by at least five radial zones, thus
suppressing numerical noise due to the reassignment of cells
to the atmosphere.
The amplitude $ A_{20}^\mathrm{E2} $ is related to the dimensionless
GW strain $ h $ in the equatorial plane at a distance $ r $ to the
source by \citep[see e.g.][]{dimmelmeier_02_b}
\begin{equation}
  h = \frac{1}{8} \sqrt{\frac{15}{\pi}} \frac{A^\mathrm{E2}_{20}}{r} =
  8.85 \times 10^{-21} \frac{A^\mathrm{E2}_{20}}{10^3 \mathrm{\ cm}}
  \frac{10 \mathrm{\ kpc}}{r}.
\end{equation}

\citet{obergaulinger_06_a} presented a comparison of two core collapse
models both simulated in GR and using the TOV potential of Case~R.
They found that the simulations with the TOV potential lead to an
error of about $ 50\% $ in the GW peak amplitude
$ |A_{20}^\mathrm{E2}|_\mathrm{max} $ (that is reached around bounce)
for model A1B3G3 despite the fact that e.g.\ the evolution of the
$ \rho_\mathrm{max} $ is quite similar in both cases. This is
confirmed by our simulations of a large and representative set of
stellar core collapse models both with a simple matter model and also
with more detailed microphysics.

\begin{figure}
  \resizebox{\hsize}{!}
  {\includegraphics{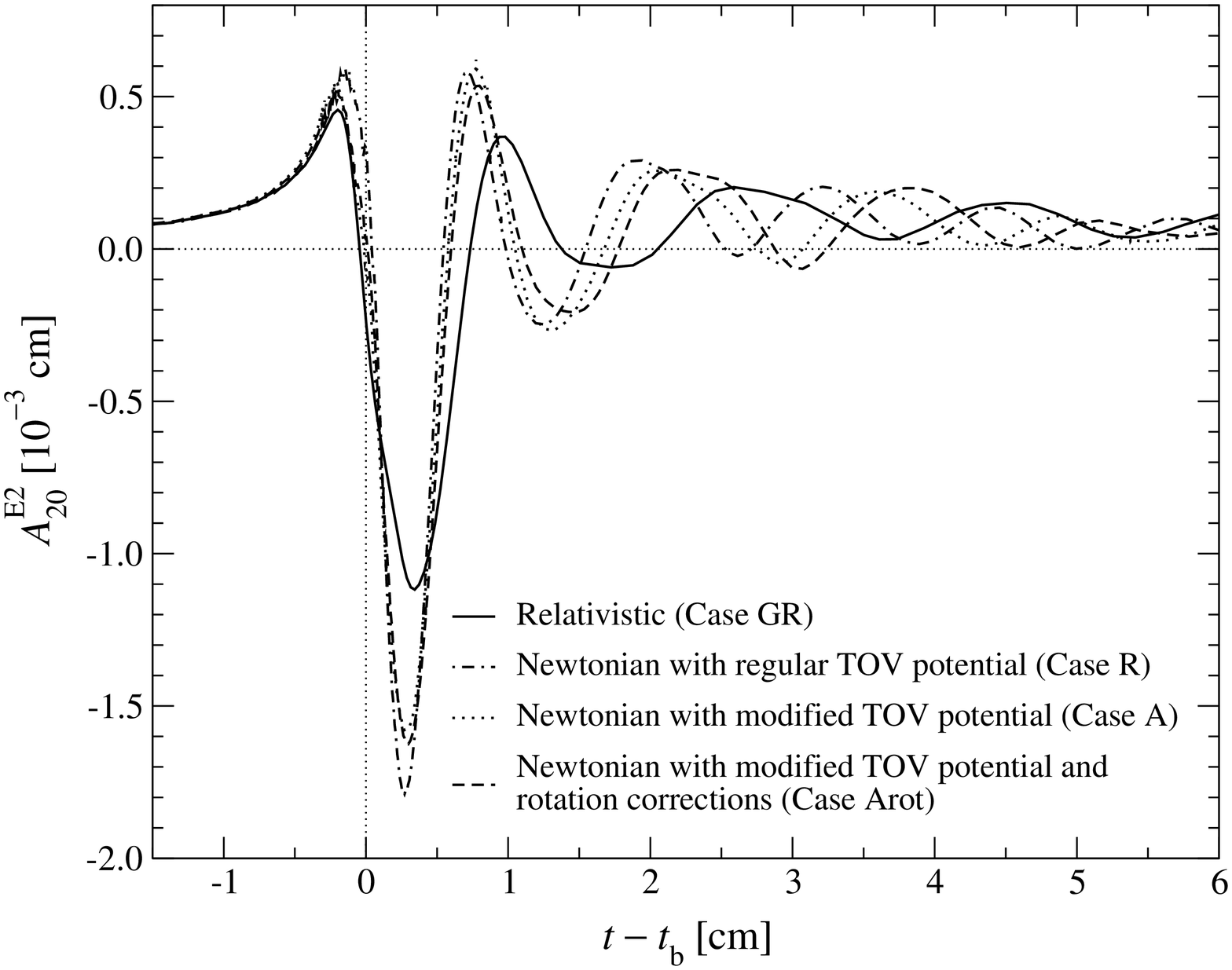}}
  \\ [1.0 em]
  \resizebox{\hsize}{!}
  {\includegraphics{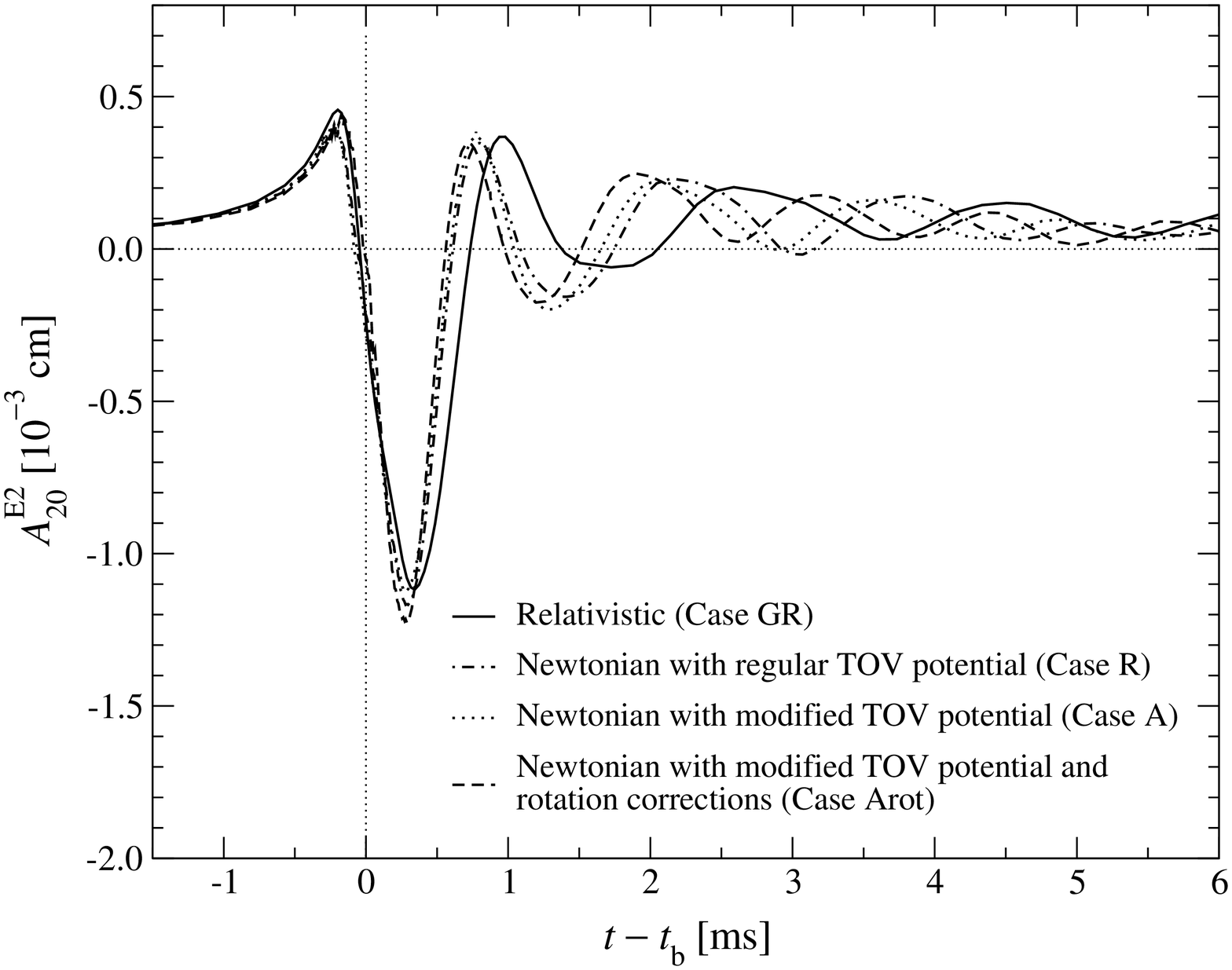}}
  \caption{Time evolution of the GW amplitude $ A_{20}^\mathrm{E2} $
    for model A1B1G1 extracted from a simulation in Case~GR (solid),
    Case~R (dashed-dotted), Case~A (dotted), and Case~Arot (dashed),
    respectively, using the time-integrated quadrupole formula. The
    top panel shows the waveforms obtained without the transformation
    to the isotropic radial coordinate for the simulations with an
    effective relativistic potential. If the transformation is
    included, the difference in the GW amplitudes compared to Case~GR
    is reduced considerably (lower panel).}
  \label{fig:waveforms}
\end{figure}

For slowly rotating models, where hydrodynamic quantities like
$ \rho_\mathrm{max} $, density profiles, etc.\ obtained using an
effective relativistic potential agree with those in GR very well, the
errors in $ |A_{20}^\mathrm{E2}|_\mathrm{max} $ (and the entire
waveform) are significant, often between $ 30\% $ and $ 50\% $. This
is exemplified by model A1B1G1, whose waveform is shown in the top
panel of Fig.~\ref{fig:waveforms}.

The quadrupole formula is derived in the Newtonian weak-field and
low-velocity limit, and has to be generalized to a GR spacetime before
a better matching to GR results can be expected. This requires a
(non-unique) choice of the radial coordinate. In the CFC metric of
Case~GR, isotropic radial coordinates are assumed. On the other hand,
as already detailed in Sect.~\ref{sec:pns_structure}, the effective
relativistic potentials are derived from the TOV solution in
Schwarzschild radial coordinates. The analogy to the discussion of the
spatial structure of the PNS suggests that consistency between GR and
Newtonian-based simulations requires the use of the same gauge and
coordinate choice for any comparison of GW amplitudes.

\begin{table*}

  \caption{Absolute value of the GW peak amplitude
    $ |A_{20}^\mathrm{E2}|_\mathrm{max} $ for all investigated core
    collapse models in Case~GR, or evolved with the effective
    relativistic potential of Case~R, A, and~Arot, respectively. For
    each potential, the relative deviation from Case~GR (in percent)
    is given in parentheses.}
  \label{tab:results_waveforms}

  \begin{center}
    \begin{tabular}{l@{\qquad}c@{\qquad}ccc@{\qquad}ccc}
      \hline \hline
      & & \multicolumn{3}{c}{no radial coordinate transformation\qquad~} &
      \multicolumn{3}{c}{transformation to isotropic radial coordinate\qquad~} \\ [0.5 em]
      Model & Case~GR & Case~R & Case~A & Case~Arot &
      Case~R & Case~A & Case~Arot \\
      \hline
      A1B1G1 &
      $ 1128 $ &
      $ 1789 ~ (+58) $ &
      $ 1581 ~ (+40) $ &
      $ 1627 ~ (+44) $ &
      $ 1227 ~ (+9)\z $ &
      $ 1170 ~ (+4)\z $ &
      $ 1127 ~ (\pm 0)\z $ \\
      A1B2G1 &
      $ 2153 $ &
      $ 3057 ~ (+42) $ &
      $ 2767 ~ (+29) $ &
      $ 2861 ~ (+33) $ &
      $ 2062 ~ (-4)\z $ &
      $ 1955 ~ (+9)\z $ &
      $ 1980 ~ (-8)\z $ \\
      A1B3G1 &
      $ 3340 $ &
      $ 4242 ~ (+27) $ &
      $ 3588 ~ (+7)\z $ &
      $ 4239 ~ (+27) $ &
      $ 2871 ~ (-14) $ &
      $ 2595 ~ (-22) $ &
      $ 2858 ~ (-15) $ \\
      A1B3G2 &
      $ 2094 $ &
      $ 2767 ~ (+32) $ &
      $ 2508 ~ (+20) $ &
      $ 2736 ~ (+31) $ &
      $ 1956 ~ (-5)\z $ &
      $ 1838 ~ (-12) $ &
      $ 1938 ~ (-15) $ \\
      A1B3G3 &
      $ \z778 $ &
      $ 1104 ~ (+42) $ &
      $ 1049 ~ (+35) $ &
      $ 1055 ~ (+36) $ &
      $ \z837 ~ (+8)\z $ &
      $ \z810 ~ (+4)\z $ &
      $ \z801 ~ (-7)\z $ \\ [0.5 em]
      A2B4G1 &
      $ \z651 $ &
      $ \z630 ~ (-3)\z $ &
      $ \z615 ~ (-6)\z $ &
      $ \z665 ~ (+2)\z $ &
      $ \z572 ~ (-12) $ &
      $ \z562 ~ (-14) $ &
      $ \z604 ~ (-7)\z $ \\ [0.5 em]
      A3B1G1 &
      $ 2456 $ &
      $ 3711 ~ (+51) $ &
      $ 3238 ~ (+32) $ &
      $ 3372 ~ (+37) $ &
      $ 2367 ~ (-4)\z $ &
      $ 2187 ~ (-11) $ &
      $ 2213 ~ (-10) $ \\
      A3B2G4 &
      $ \z598 $ &
      $ \z791 ~ (+32) $ &
      $ \z762 ~ (+27) $ &
      $ \z779 ~ (+30) $ &
      $ \z623 ~ (+4)\z $ &
      $ \z606 ~ (+1)\z $ &
      $ \z608 ~ (+2)\z $ \\
      A3B4G2 &
      $ \z842 $ &
      $ \z802 ~ (-5)\z   $ &
      $ \z775 ~ (-8)\z   $ &
      $ \z900 ~ (+7)\z   $ &
      $ \z731 ~ (-13)  $ &
      $ \z705 ~ (-16)  $ &
      $ \z790 ~ (-6)\z $ \\ [0.5 em]
      A4B1G1 &
      $ 3269 $ &
      $ 4610 ~ (+41) $ &
      $ 3746 ~ (+15) $ &
      $ 4384 ~ (+34) $ &
      $ 2836 ~ (-13) $ &
      $ 2485 ~ (-24) $ &
      $ 2652 ~ (-19) $ \\
      A4B1G2 &
      $ 2745 $ &
      $ 3593 ~ (+31) $ &
      $ 3041 ~ (+11) $ &
      $ 3550 ~ (+29) $ &
      $ 2382 ~ (-13) $ &
      $ 2103 ~ (-23) $ &
      $ 2284 ~ (-17) $ \\
      A4B2G2 &
      $ 4490 $ &
      $ 5067 ~ (+13)   $ &
      $ 4393 ~ (+2)\z  $ &
      $ 5704 ~ (+27)   $ &
      $ 3444 ~ (-23) $ &
      $ 3084 ~ (-31) $ &
      $ 3581 ~ (-20) $ \\
      A4B2G3 &
      $ 3093 $ &
      $ 3546 ~ (+15) $ &
      $ 3032 ~ (-2)\z $ &
      $ 3823 ~ (+24) $ &
      $ 2597 ~ (-16) $ &
      $ 2407 ~ (-22) $ &
      $ 2606 ~ (-16) $ \\
      A4B5G4 &
      $ 2466 $ &
      $ 2754 ~ (+12)   $ &
      $ 2725 ~ (+11)   $ &
      $ 2669 ~ (+8)  \z $ &
      $ 2505 ~ (+2)\z $ &
      $ 2479 ~ (+1)\z $ &
      $ 2345 ~ (-5)\z $ \\
      A4B5G5 &
      $ 3383 $ &
      $ 3766 ~ (+11) $ &
      $ 3744 ~ (+11) $ &
      $ 3614 ~ (+7)\z $ &
      $ 3476 ~ (+3)\z $ &
      $ 3457 ~ (+2)\z $ &
      $ 3258 ~ (-4)\z $ \\ [0.5 em]
      s20A1B1 &
      $ \z\z 69 $ &
      $ \z 108 ~ (+57) $ &
      $ \z 101 ~ (+46) $ &
      $ \zz 89 ~ (+29) $ &
      --- &
      --- &
      $ \zz 68 ~ (-1)\z $ \\
      s20A2B4 &
      $ \z 339 $ &
      $ \z 330 ~ (-3)\z $ &
      $ \z 310 ~ (-9)\z $ &
      $ \z 353 ~ (+4)\z $ &
      --- &
      --- &
      $ \z321 ~ (-5)\z $ \\
      \hline \hline
    \end{tabular}
  \end{center}
\end{table*}

We therefore apply the transformation from the Schwarzschild radial
coordinate to the isotropic radial coordinate (see Eq.~(B.8) in
Paper~I) when extracting GW signals via the quadrupole formula. This
transformation effectively reduces the magnitude of the factor $ r^3 $
in Eq.~(\ref{eq:quadrupole_formula}) and thus leads to a considerably
lower amplitude, while the waveform shape itself remains essentially
unchanged. In many cases, the coordinate transformation significantly
improves the agreement with the GR results: For the slow and almost
uniform rotator A1B1G1 (see lower panel of Fig.~\ref{fig:waveforms}),
$ A_{20}^\mathrm{E2} $ extracted from a simulation with any of the
effective relativistic potentials now deviates by less than $ 10\% $
from the GR simulation, as opposed to more than $ 40\% $ without the
coordinate change. For the microphysical model s20A1B1, which has the
same initial density profile and rotation rate, the effect is also
dramatic: The deviation is reduced from $ 29\% $ to $ 1\% $ for
Case~Arot.

We observe a similar improvement for most other investigated models
with slow and weakly differential rotation, i.e.\ where
$ T / |W| \lesssim 0.1 $ during the whole evolution: Table~\ref{tab:results_waveforms}
shows that for the potentials $ \Phi_\mathrm{R} $ and $ \Phi_\mathrm{Arot} $
the deviation from GR is reduced in such cases (all the $A1$ and  $B1$ models),
and there are only three exceptions for potential $ \Phi_\mathrm{A} $.
Even for fast rotators (see e.g.\ the
B3, B4, and B5 models in Table~\ref{tab:results_waveforms}), we
observe no errors in the bounce signal larger than $ 20\% $
(Case~Arot) and $ 24\% $ (Case~R), respectively, when the coordinate
transformation is applied. For more rapid rotators, Case~A is an
exception as the transformation actually degrades the quality of the
results. This is to be expected, as that potential often leads to a
core density (i.e.\ compactness) at bounce that is significantly below
the value obtained with potential $ \Phi_\mathrm{R} $ (which always
gives larger densities than potential $ \Phi_\mathrm{A} $) and
potential $ \Phi_\mathrm{Arot} $ (which includes corrections due to
rotation).

In summary we can infer that the combination of the potential
$ \Phi_\mathrm{Arot} $ and the radial coordinate transformation in the
GW extraction based on the quadrupole formula delivers waveforms that
are very accurate when compared to the ones from GR simulations for a
wide range of rotation rates and profiles of collapsing compact
objects.


\section{Summary and conclusions}
\label{sec:summary}

We have compared pseudo-Newtonian simulations of rotational stellar
core collapse with three different effective relativistic
gravitational potentials to reference simulations with the conformal
flatness approximation of general relativity. This has been done using
either a simplified EoS or a microphysical treatment of matter. We
have been able to improve the effective relativistic potential
approach used by \citet{marek_05_a} and \citet{obergaulinger_06_a} both
in the regime of slow and rapid rotation. Our tests also allow us to
assess the applicability of the different old potentials as well as
the new improvement in these two regimes.

In the case of slow and moderate rotation, all potentials give good
results for the dynamical evolution. This implies that the generic
scenario of core collapse in massive stars can probably be treated
without including any rotational corrections like implemented in the
new potential $ \Phi_\mathrm{Arot} $, because the cores of these stars
are not expected to spin rapidly \citep{heger_00_a}. Moreover, the
multiple bounce scenario, which is most problematic for the
rotationally uncorrected potentials $ \Phi_\mathrm{A} $ and
$ \Phi_\mathrm{R} $, is suppressed once deleptonization is taken into
account. However, we have demonstrated that the gravitational wave
signal can still be significantly improved also in these cases by
applying a simple coordinate transformation when utilizing the
Newtonian quadrupole formula. Our new effective relativistic potential
now yields wave amplitudes which are within $ 20\% $ of the values
found in relativistic simulations, as opposed to errors of more than
$ 50\% $ reported by \citet{obergaulinger_06_a}. In particular, we
observe excellent agreement for the most slowly rotating models
considered, i.e.\ the ones closest to the predictions of
\citet{heger_00_a} in terms of rotation rates. The best overall match
with relativistic gravitational wave signals has been obtained with
$ \Phi_\mathrm{Arot} $, which is identical to $ \Phi_\mathrm{A} $ in
spherical symmetry. As in this limit $ \Phi_\mathrm{A} $ is superior
to the original effective relativistic potential $ \Phi_\mathrm{R} $
\citep{marek_05_a}, the potential $ \Phi_\mathrm{Arot} $ is already a
very attractive choice in the regime of slow rotation.

For strong rotation we have studied more demanding test problems for
our potentials than considered in Paper~I. In order to assess in
general the usefulness of the effective relativistic potential
approach for prospective simulations of rotational instabilities in a
proto-neutron star or a cold neutron star, or for the
accretion-induced collapse scenario, we have investigated models of a
rotating neutron star, and rotational stellar core collapse with high
rotation rates and an extremely differential angular momentum
distribution. We have included very rapidly rotating core collapse
models for which $ T / |W| \approx 0.4 $ is reached, and have also
discussed equilibrium models of a neutron star with strong
differential rotation and axis ratios down to
$ r_\mathrm{p} / r_\mathrm{e} = 0.05 $. The rotational parameters of
these models were chosen such as to also provide a reasonable
description of the accretion-induced collapse scenario, which has
however not been directly simulated in this work.

As in Paper~I we confirm that the accuracy of Cases~R and~A declines
for rapidly rotating configurations. For instance, these potentials
lead to a gross underestimation of the maximum density at bounce for
core collapse models with a single centrifugal bounce, such as model
A4B5G5, where the deviation to the general relativistic results is as
large as $ 27\% $. These `old' potentials also cannot reproduce the
structure of a rapidly rotating neutron star, although they still
provide a clear improvement over the purely Newtonian treatment. By
including rotational corrections in our new potential
$ \Phi_\mathrm{Arot} $, we have been able to obtain much closer
agreement with the general relativistic results in many core collapse
simulations as well as for rapidly rotating neutron star models in
equilibrium. In Case~Arot the maximum density at bounce, for example,
now always lies within $ 10\% $ of the correct value, and radial
profiles of strongly rotating and toroidally deformed models also
agree very well with the ones from general relativistic simulations.
Thus, in contrast to potentials $ \Phi_\mathrm{R} $ and
$ \Phi_\mathrm{A} $, we infer that for all practical purposes the new
potential $ \Phi_\mathrm{Arot} $ works equally well from the
non-rotating limit up to very rapidly and differentially rotating
configurations.

However, a computation of neutron star eigenfrequencies has revealed
large errors of up to $ 50\% $ for the frequencies of the $ F $-mode
and its harmonics, which implies that even in spherical symmetry some
of the dynamic properties are not captured correctly by any of the
effective relativistic potentials investigated. These shortcomings in
accurately approximating the correct pulsation frequencies reveals an
inherent flaw of the effective relativistic potential approach.

Our findings have important implications for the range of
applicability of an effective relativistic potential in an otherwise
Newtonian code. The rotation rate of our most rapidly rotating models
of (proto-)neutron stars exceed values of $ T / |W| \approx 0.3 $, which
is the expected extreme upper limit for rotational supernova core
collapse \citep{ott_07_b, dimmelmeier_08_a} or accretion-induced
collapse \citep{dessart_06_a}, and lies well beyond the Kepler limit for
uniformly rotating neutron stars\footnote{Such rotation rates are also
above the classical thresholds for the growth of triaxial
instabilities on a secular time scale ($ T / |W| \geq 0.14 $)
and on a dynamical time scale ($ T / |W|\geq 0.27 $).}. While
the potentials $ \Phi_\mathrm{R} $ and $ \Phi_\mathrm{A} $, which lack
rotational corrections, are already quite unreliable in this regime,
such high rotation rates pose no problem for the new potential
$ \Phi_\mathrm{Arot} $. Thus, our new approach is clearly the
preferred choice for modeling rapidly rotating compact objects in the
framework of pseudo-Newtonian simulations.


\begin{acknowledgements}
  This research has been supported by the DFG (SFB/Transregio 7 and
  SFB 375) and by the DAAD and IKY (IKYDA German--Greek research
travel grant). H.~D.\ is a Marie Curie Intra-European Fellow within the
  6th European Community Framework Programme (IEF 040464). It is a
  pleasure to thank C.~D.~Ott, A.~Marek, and H.-T.~Janka for their
  contributions related to the improved microphysics. We also thank
  the referee for his suggestions to improve our manuscript.
\end{acknowledgements}



\begin{thebibliography}{99}

\bibitem[\protect\citeauthoryear{Banyuls et~al.}{1997}]{banyuls_97_a}
  Banyuls, F., Font, J.~A., Ib{\'a}{\~n}ez, J.~M., Mart\'{\i}, J.~M., \& Miralles, J.~A. 1997,
  \apj, 476, 221

\bibitem[\protect\citeauthoryear{Blanchet et~al.}{1990}]{blanchet_90_a}
  Blanchet, L., Damour, T., \& Sch{\"a}fer, G. 1990,
  \mnras, 242, 289

\bibitem[\protect\citeauthoryear{Burrows et~al.}{2006}]{burrows_06_a}
  Burrows, A., Livne, E., Dessart, L., Ott, C.~D., \& Murphy, J. 2006
  \apj, 640, 878

\bibitem[\protect\citeauthoryear{Burrows et~al.}{2007}]{burrows_07_a}
  Burrows, A., Livne, E., Dessart, L., Ott, C.~D., \& Murphy, J. 2007
  \apj, 655, 416

\bibitem[\protect\citeauthoryear{Cantiello et~al.}{2007}]{cantiello_07_a}
  Cantiello, M., Yoon, S.-C., Langer, N., \& Livio, M. 2007
  \aap, 465, L29

\bibitem[\protect\citeauthoryear{Centrella et~al.}{2001}]{centrella_01_a}
  Centrella, J.~M., New, K.~C.~B., Lowe, L.~L., \& Brown, J.~D. 2001,
  \apjl, 550, L193

\bibitem[\protect\citeauthoryear{Cerd\'a-Dur\'an et~al.}{2005}]{cerda_05_a}
  Cerd\'a-Dur\'an, P., Faye, G., Dimmelmeier, H., et~al. 2005,
  \aap, 439, 1033

\bibitem[\protect\citeauthoryear{Dessart et~al.}{2007a}]{dessart_07_a}
  Dessart, L., Burrows, A., Livne, E., \& Ott, C.~D. 2007a,
  \apj, 669, 585
  
\bibitem[\protect\citeauthoryear{Dessart et~al.}{2007b}]{dessart_07_b}
  Dessart, L., Burrows, A., Livne, E., \& Ott, C.~D. 2007b,
  \apjl, 673, L43

\bibitem[\protect\citeauthoryear{Dessart et~al.}{2006}]{dessart_06_a}
  Dessart, L., Burrows, A., Ott, C.~D., et~al. 2006,
  \apj, 644, 1063

\bibitem[\protect\citeauthoryear{Dimmelmeier et~al.}{2006}]{dimmelmeier_06_a}
  Dimmelmeier, H., {Cerd{\'a}-Dur{\'a}n}, P., Marek, A., \& Faye, G. 2006,
  in AIP Conference Series, Vol. 861, Albert Einstein Century International
  Conference, ed. J.-M. Alimi \& A.~F{\"u}zfa
  (Melville, USA: American Institute of Physics), 600

\bibitem[\protect\citeauthoryear{Dimmelmeier et~al.}{2002a}]{dimmelmeier_02_a}
  Dimmelmeier, H., Font, J.~A., \& M{\"u}ller, E. 2002a,
  \aap, 388, 917

\bibitem[\protect\citeauthoryear{Dimmelmeier et~al.}{2002b}]{dimmelmeier_02_b}
  Dimmelmeier, H., Font, J.~A., \& M{\"u}ller, E. 2002b,
  \aap, 393, 523

\bibitem[\protect\citeauthoryear{Dimmelmeier et~al.}{2005}]{dimmelmeier_05_a}
  Dimmelmeier, H., Novak, J., Font, J.~A., Ib{\'a}{\~n}ez, J.~M., \& M\"uller, E. 2005,
  \prd, 71, 064023

\bibitem[\protect\citeauthoryear{Dimmelmeier et~al.}{2007}]{dimmelmeier_07_a}
  Dimmelmeier, H., Ott, C.~D., Janka, H.-T., Marek, A., \& M\"uller, E. 2007,
  Phys. Rev. Lett., 98, 251101

\bibitem[\protect\citeauthoryear{Dimmelmeier et~al.}{2008}]{dimmelmeier_08_a}
  Dimmelmeier, H., Ott, C.~D., Marek, A., \& Janka, H.-T. 2008, in preparation

\bibitem[\protect\citeauthoryear{Finn}{1989}]{finn_89_a}
  Finn, L.~S. 1989,
  in Evans, C.~R. and Finn, L.~S. and Hobill, D.~W., eds.,
  Frontiers in Numerical Relativity
  (Cambridge, UK: Cambridge University Press), 126
  
\bibitem[\protect\citeauthoryear{Fryer \& Heger}{2005}]{fryer_05_a}
  Fryer, C.~L. \& Heger, A. 2005
  \apj, 623, 302

\bibitem[\protect\citeauthoryear{Heger et~al.}{2000}]{heger_00_a}
  Heger, A., Langer, N., \& Woosley, S.~E. 2000,
  \apj, 528, 368

\bibitem[\protect\citeauthoryear{Heger et~al.}{2005}]{heger_05_a}
  Heger, A., Woosley, S.~E., \& Spruit, H.~C. 2005
  \apj, 626, 350

\bibitem[\protect\citeauthoryear{Isenberg}{1978}]{isenberg_78_a}
  Isenberg, J.~A. 1978,
  University of Maryland Preprint, preprint \texttt{[arXiv:gr-qc/0702113]}

\bibitem[\protect\citeauthoryear{Janka et~al.}{1993}]{janka_93_a}
  Janka, H.-T., Zwerger, T., \& M\"onchmeyer, R. 1993,
  \aap, 268, 360

\bibitem[\protect\citeauthoryear{Komatsu et~al.}{1989}]{komatsu_89_a}
  Komatsu, H., Eriguchi, Y., \& Hachisu, I. 1989,
  \mnras, 237, 355

\bibitem[\protect\citeauthoryear{Langanke \& {Mart{\'{\i}}nez-Pinedo}}{2000}]{langanke_00_a}
  Langanke, K. \& Mart{\'{\i}}nez-Pinedo, G. 2000,
  Nucl. Phys. A, 673, 481

\bibitem[\protect\citeauthoryear{Liebend{\"o}rfer}{2005}]{liebendoerfer_05_b}
  Liebend{\"o}rfer, M. 2005,
  \apj, 633, 1042

\bibitem[\protect\citeauthoryear{Liebend\"orfer et~al.}{2005}]{liebendoerfer_05_a}
  Liebend\"orfer, M., Rampp, M., Janka, H.-T., \& Mezzacappa, A. 2005,
  \apj, 620, 840

\bibitem[\protect\citeauthoryear{Lin et~al.}{2006}]{lin_06_a}
  Lin, L.-M., Cheng, K.~S., Chu, M.-C., \& Suen, W.-M. 2006,
  \apj, 639, 382

\bibitem[\protect\citeauthoryear{MacFadyen \& Woosley}{1999}]{macfadyen_99_a}
  MacFadyen, A.~I. \& Woosley, S.~E. 1999,
  \apj, 524, 262

\bibitem[\protect\citeauthoryear{Marek et~al.}{2006}]{marek_06_a}
  Marek, A., Dimmelmeier, H., Janka, H.-T., M{\"u}ller, E., \& Buras, R. 2006,
  \aap, 445, 273

\bibitem[\protect\citeauthoryear{Marek et~al.}{2005}]{marek_05_a}
  Marek, A., Janka, H.-T., Buras, R., Liebend{\"o}rfer, M., \& Rampp, M. 2005,
  \aap, 443, 201

\bibitem[\protect\citeauthoryear{M\"onchmeyer et~al.}{1991}]{moenchmeyer_91_a}
  M\"onchmeyer, R., Sch\"afer, G., M\"uller, E., \& Kates, R.~E. 1991,
  \aap, 246, 417

\bibitem[\protect\citeauthoryear{Obergaulinger et~al.}{2006}]{obergaulinger_06_a}
  Obergaulinger, M., Aloy, M.~A., Dimmelmeier, H., \& M{\"u}ller, E. 2006,
  \aap, 457, 209

\bibitem[\protect\citeauthoryear{Ott et~al.}{2007a}]{ott_07_a}
  Ott, C.~D., Dimmelmeier, H., Marek, A., et~al. 2007a,
  \prl, 98, 261101

\bibitem[\protect\citeauthoryear{Ott et~al.}{2007b}]{ott_07_b}
  Ott, C.~D., Dimmelmeier, H., Marek, A., et~al. 2007b,
  Class. Quantum Grav., 24, 139

\bibitem[\protect\citeauthoryear{Ott et~al.}{2005}]{ott_05_a}
  Ott, C.~D., Ou, S., Tohline, J.~E., \& Burrows, A. 2005,
  \apjl, 625, L119

\bibitem[\protect\citeauthoryear{Ou \& Tohline}{2006}]{ou_06_a}
  Ou, S. \& Tohline, J.~E. 2006,
  \apj, 651, 1068

\bibitem[\protect\citeauthoryear{Ou et~al.}{2004}]{ou_04_a}
  Ou, S., Tohline, J.~E., \& Lindblom, L. 2004,
  \apj, 617, 490

\bibitem[\protect\citeauthoryear{Rampp \& Janka}{2002}]{rampp_02_a}
  Rampp, M. \& Janka, H.-T. 2002,
  \aap, 396, 361

\bibitem[\protect\citeauthoryear{Shen et~al.}{1998}]{shen_98_a}
  Shen, H., Toki, H., Oyamatsu, K., \& Sumiyoshi, K. 1998,
  Prog. Theor. Phys., 100, 1013

\bibitem[\protect\citeauthoryear{Shibata et~al.}{2002}]{shibata_02_a}
  Shibata, M., Karino, S., \& Eriguchi, Y. 2002,
  \mnras, 334, L27

\bibitem[\protect\citeauthoryear{Shibata et~al.}{2003}]{shibata_03_a}
  Shibata, M., Karino, S., \& Eriguchi, Y. 2003,
  \mnras, 343, 619

\bibitem[\protect\citeauthoryear{Shibata \& Sekiguchi}{2004}]{shibata_04_a}
  Shibata, M. \& Sekiguchi, Y.-I. 2004,
  \prd, 69, 084024

\bibitem[\protect\citeauthoryear{Shibata \& Sekiguchi}{2005}]{shibata_05_b}
  Shibata, M. \& Sekiguchi, Y.-I. 2005,
  \prd, 71, 024014

\bibitem[\protect\citeauthoryear{Straumann}{2004}]{straumann_04_a}
  Straumann, N. 2004, General Relativity with Applications to Astrophysics
  (Berlin, Germany: Springer)

\bibitem[\protect\citeauthoryear{van Riper}{1979}]{vanriper_79_a}
  van Riper, K.~A. 1979,
  \apj, 232, 558

\bibitem[\protect\citeauthoryear{Weinberg \& Quataert}{2008}]{weinberg_08_a}
  Weinberg, N.~N. \& Quataert, E. 2008
  \mnras, accepted, preprint \texttt{[arXiv:0802.1522v2]}

\bibitem[\protect\citeauthoryear{Wilson et~al.}{1996}]{wilson_96_a}
  Wilson, J.~R., Mathews, G.~J., \& Marronetti, P. 1996,
  \prd, 54, 1317

\bibitem[\protect\citeauthoryear{Woosley \& Heger}{2006}]{woosley_06_a}
  Woosley, S.~E. \& Heger, A. 2006
  \apj, 637, 914

\bibitem[\protect\citeauthoryear{Woosley et~al.}{2002}]{woosley_02_a}
  Woosley, S.~E., Heger, A., \& Weaver, T.~A. 2002,
  Rev. Mod. Phys., 74, 1015

\bibitem[\protect\citeauthoryear{Yoon \& Langer}{2005}]{yoon_05_a}
  Yoon, S.-C. \& Langer, N. 2005
  \aap, 435, 967

\bibitem[\protect\citeauthoryear{Zwerger \& M\"uller}{1997}]{zwerger_97_a}
  Zwerger, T. \& M\"uller, E. 1997,
  \aap, 320, 209

\end{thebibliography}

\end{document}